\documentclass[11pt]{article}

\PassOptionsToPackage{table}{xcolor}
\usepackage[preprint]{acl}

\usepackage{times}
\usepackage{latexsym}
\usepackage[normalem]{ulem}

\usepackage[utf8]{inputenc}

\usepackage{microtype}

\usepackage{inconsolata}

\usepackage{placeins}

\usepackage{tikz}
\usepackage{amsmath}
\usepackage{filecontents}
\usepackage[flushleft]{threeparttable}
\usepackage{amsmath,amsfonts}

\usepackage{graphicx}
\usepackage{textcomp}

 \usepackage{url}

\usepackage{lipsum,tabularx}
\usepackage{pythonhighlight}
\usepackage{multicol}
\usepackage{multirow}
\usepackage{amssymb}

\usepackage{colortbl}
\usepackage{booktabs}
\usepackage{setspace}

\hypersetup{
  linkcolor={blue!70!black},
  citecolor={red!70!black},
  urlcolor={blue!70!black}
}
\def\Snospace~{\S{}}

\usepackage[T1]{fontenc}

\usepackage{algorithm}
\usepackage{algorithmicx}
\usepackage[noend]{algpseudocode}
\usepackage{balance}
\usepackage{enumitem}

\usepackage{bm}
\usepackage{fp}
\usepackage{siunitx}
\usepackage{amsthm}

\usepackage[labelfont=bf,font=small,skip=5pt]{caption}
\usepackage{subcaption}

\usepackage{comment}

\usepackage{tikz}

\usepackage{xspace}

\newcommand{\boxbeg}{
  \vspace{2px}
  \noindent\begin{tabular}{|l|}\hline
    \begin{minipage}{3.2in}
      \vspace{2px}
      \noindent
      }

      \newcommand{\boxend}{
      \vspace{2px}
    \end{minipage} \\ \hline
  \end{tabular}
  \vspace{-10pt}
}

\usepackage{algpseudocode}
\usepackage[htt]{hyphenat}

\usepackage{wrapfig}

\usepackage[scaled=.7]{beramono}
\usepackage{pifont}
\usepackage{mdframed}
\usepackage[most]{tcolorbox}
\newtcolorbox{promptbox}[1][]{
  enhanced,
  breakable,
  title=#1,
  colback=gray!5, 
  colframe=gray!60!black,
  colbacktitle=gray!80,
  coltitle=white,
  boxrule=0.7pt,
  arc=2mm,
  left=4mm,right=4mm,top=2mm,bottom=2mm,
  fonttitle=\bfseries,
}

\newcounter{finding}

\newcommand{\nop}[1]{}
\newcommand{\sys}{\textsc{CAPE}\xspace} 

\title{Out of Sight: Compression-Aware Content Protection against Agentic Crawlers}

\author{
  Xuefei Wang
  \\
  Beihang University
  \\
  \texttt{xuefeiw@buaa.edu.cn}
}

\begin{document}

\maketitle

\begin{abstract}
The rise of LLM-based agents with reasoning, summarization, and memory capabilities has created a new threat surface for online content that conventional defenses fail to address.
Existing defenses like access controls can be circumvented by agents mimicking ordinary browsers, and injection-based defenses often degrade human readability.
In this paper, we revisit the agent pipeline and identify context compression, which agents routinely invoke to fit context budgets, as a critical yet overlooked defense layer.
We propose \sys, a framework that protects high-value textual content by injecting invisible perturbations without changing its human-visible surface form, thereby inducing severe information loss during agent compression.
\sys extracts disruptive seed perturbations from an accessible surrogate compressor, then adapts them to query-only target compressors through prior-guided evolution and preference-calibrated candidate prioritization, achieving effective protection under a low query budget.
Experiments on three content types and four compression settings show that \sys improves information loss by up to 75.8\% over the strongest baseline while keeping protected content visually indistinguishable from originals.
\sys also transfers to real-world settings, including the LangGraph agent workflow and GitHub Copilot, highlighting its generality and practical value.
This paper aims to reveal context compression as a new defense layer, promoting content protection research in the agent era.
\end{abstract}

\section{Introduction}\label{sec:intro}

Unauthorized scraping has long been a critical threat to online knowledge and digital assets, as highlighted by lawsuits such as The New York Times against Microsoft~\cite{nyt2023complaint} and Reddit against Anthropic~\cite{reddit2025anthropic}.
The widespread deployment of large language model (LLM) agents in web environments has severely exacerbated this issue.
Unlike traditional scrapers that store pages verbatim, agent-driven crawlers possess strong reasoning, summarization, and memory capabilities.
They actively chain retrieval with context compression and memory writing, reusing condensed information across downstream tasks~\cite{jiang2023llmlingua,pan-etal-2024-llmlingua,chevalier2023adapting,ge2023context}.
In addition, recent studies reveal that automated traffic now accounts for nearly half of all internet activity, with certain AI scraping bots experiencing up to a 300\% surge in requests within a single year~\cite{imperva2024badbot,cloudflare2025crawlers}.
These trends underscore the urgent need for content protection methods tailored to the agentic era.

Despite the various methods proposed to defend against and prevent unauthorized crawling, they exhibit clear limitations when facing modern agentic crawlers.
First, access control mechanisms such as \texttt{robots.txt}, Cloudflare's one-click block~\cite{cloudflare2024aiblock}, and standardized licensing protocols~\cite{w3c2024tdmrep,iptc2023datamining} rely on crawler identification and voluntary compliance, both of which agents easily circumvent by issuing requests from regular browser sessions~\cite{kim2025scrapers,zhong2025web}.
Since these mechanisms act before content is delivered, they are rendered powerless once the agent successfully slips through and retrieves the page.
In addition, agent-targeted attacks such as prompt injection and jailbreaking~\cite{zou2023universal,liu2023autodan,liu2024formalizing,debenedetti2024agentdojo,wei2023jailbroken} primarily mislead agent outputs rather than protect the underlying content.
Their disruptive perturbations will also raise compliance issues and degrade user readability, limiting practical value as a content protection mechanism.

To fill this gap, we revisit the structure of agent pipelines and identify context compression as a critical chokepoint~\cite{jiang2023llmlingua,pan-etal-2024-llmlingua,chevalier2023adapting,packer2023memgpt}.
Since compression is the inevitable interface processing raw text to fit context budgets, intervening at this stage provides an orthogonal layer of defense.
Rather than competing with existing access controls, it serves as a complementary fallback that activates when perimeter defenses are bypassed.
Based on this insight, we propose \sys, a framework that injects invisible perturbations without changing the human-visible surface form into protected documents, causing severe information loss during agent compression.
\sys solves this constrained optimization problem through three stages:
\ding{182} \textit{Structural prior discovery} stage leverages an accessible compressor to identify degradation-maximizing seed perturbations and builds a seed corpus with their structural priors and scores.
\ding{183} \textit{Prior-guided evolutionary adaptation} stage applies prior-guided recombination, mutation, and selection on the seeds to search target-adapted candidates.
\ding{184} \textit{Preference-calibrated query selection} stage learns degradation direction from compression feedback to prioritize high-value candidates and allocate limited queries, enabling low-cost target adaptation.
Experiments on three high-value content types (long-form text, code, and dialogue histories) show that \sys improves information degradation by up to 75.8\% over the strongest baseline, while keeping protected text visually identical to the originals (human-visible difference: 1.4\%).
The protection also transfers to real-world agentic pipelines (LangGraph and GitHub Copilot), inducing up to a 59.7\% accuracy drop in downstream tasks.

Our contributions are as follows:
\ding{182} We are the first to identify context compression as a defense layer for content protection in the agent era, establishing a content-layer defense that complements existing access-control methods.
\ding{183} We formulate perturbation design as a multi-objective optimization constraining human-visible surface difference while degrading compression quality, and propose \sys, a unified framework that combines distributional prior discovery, prior-guided evolution, and preference-calibrated query selection.
\ding{184} Systematic experiments on three content types and four target settings show that \sys improves information loss by up to 75.8\% over the strongest baseline, while the protected input remains readable as the original content.
\sys also transfers to real-world agents, producing a 59.7\% accuracy drop in LangGraph and GitHub Copilot workflows.
\ding{185} We publicly release the prototype of \sys and necessary evaluation materials to support reproducibility\footnote{\url{https://anonymous.4open.science/r/CAPE-580C}}.

\section{Background and Related Work} \label{sec:6_related}

\begin{figure}[t]
    \centering
    \includegraphics[width=0.85\columnwidth]{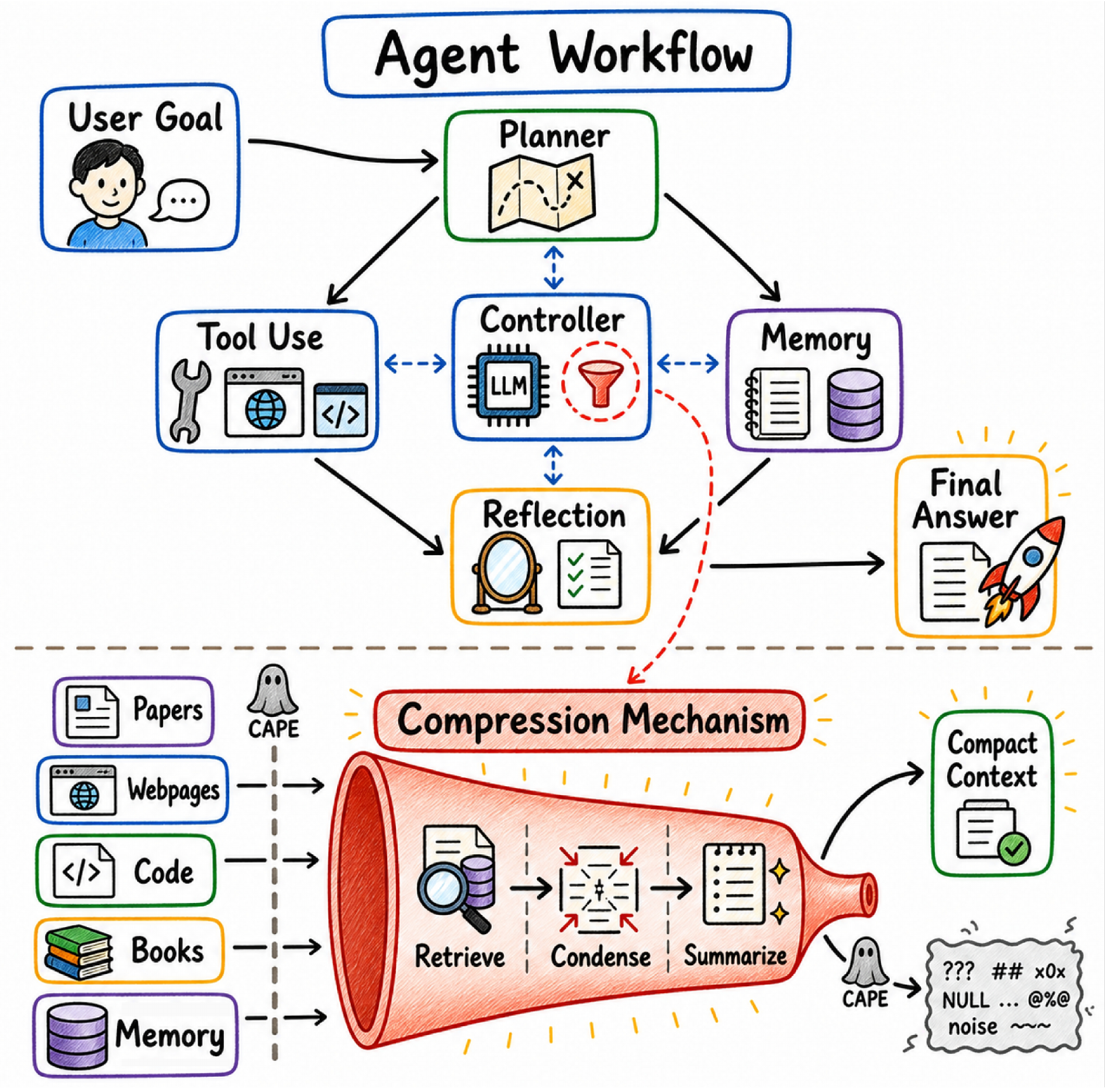}
    \caption{General agent workflow with compression mechanism.}
    \label{fig:agent-workflow}
    \vspace{-5mm}
\end{figure}

\noindent
\textbf{Agent Workflows.}
Modern LLM-based agents translate a user goal into a final answer through a multi-stage, closed-loop workflow~\cite{xi2025rise,sumers2024cognitive,wang2024survey}.
As illustrated in~\autoref{fig:agent-workflow}, this pipeline can be decomposed into five interacting stages.
\ding{182} \textit{Planner} decomposes the goal into actionable sub-tasks~\cite{chakraborty2026t1}.
\ding{183} \textit{Tool Use} stage uses browsers, code interpreters, and APIs to gather external evidence such as web pages, papers, and code~\cite{shi2025prompt}.
\ding{184} \textit{Memory} stores interaction history and retrieved knowledge and evidence~\cite{park2023generative,packer2023memgpt}.
\ding{185} \textit{Controller} (the backbone LLM) integrates these signals and drives intermediate reasoning.
\ding{186} \textit{Reflection} iteratively verifies and refines outputs before emitting the final answer.
Crucially, when the agent invokes external tools or accesses memory to gather evidence like web pages or code, the retrieved raw data routinely exceeds context budgets.
Therefore, the workflow implements an embedded \textit{context compression} module to condense and summarize external content before reasoning.
Researchers have proposed various compression approaches~\cite{pan-etal-2024-llmlingua,ge2023context,chevalier2023adapting,gilbert2023semantic,kim2024compressed}, among which one of the mainstream designs is LLM-driven compression.
It leverages the backbone model or a dedicated assistant model to rewrite retrieved documents, web pages, and interaction histories into compact natural language abstracts~\cite{xu2023recomp,yoon2024compact,lee2024readagent,zhong2024memorybank}.

\noindent
\textbf{Textual Content Protection.}
Traditional defenses protect text via publisher-level opt-outs (e.g., \texttt{robots.txt}, standard protocols~\cite{w3c2024tdmrep,iptc2023datamining}) and network-level traffic filters (e.g., CAPTCHAs, AI crawler blocks~\cite{imperva2024badbot,cloudflare2024aiblock,cloudflare2025crawlers,zhong2025web}).
Crucially, these mechanisms act before content delivery, rendering them defenseless once an agentic crawler successfully retrieves the rendered page, as modern agents easily mimic human navigation to bypass filters and selectively ignore compliance rules~\cite{kim2025scrapers}.
Furthermore, defenses operating elsewhere in the content lifecycle, such as post-hoc watermarking~\cite{lau2024waterfall,wu2024enhancing} or training-time poisoning~\cite{java2025towards,zhao2025data}, cannot intervene during inference-time ingestion.
This highlights an urgent need for an inference-time, readability-preserving content protection method tailored for the agentic era.

\noindent
\textbf{Attacks on LLM Agents.}
Existing attacks hijack agent pipelines via indirect prompt injections~\cite{greshake2023not,debenedetti2024agentdojo,liu2024formalizing,shi2025prompt}, memory poisoning~\cite{chen2024agentpoison}, and adversarial suffixes~\cite{zou2023universal,liu2023autodan,paulus2024advprompter,jia2024improved}.
Researchers have also designed subtle perturbations, such as typo-based attacks~\cite{cho2024typos} and adversarial tokenization~\cite{geh2025adversarial,zhuo2025ability}, with recent work specifically exploiting context compression to steer downstream behavior~\cite{liu2025compressionattack}.
While these efforts use perturbations to hijack agents into unauthorized actions, \sys designs invisible perturbations solely to corrupt the semantic integrity of compression outputs, achieving content protection without attempting to control agent behavior.
\section{Problem Setting and Threat Model}\label{sec:3_Problem}




\subsection{Threat Model}
\label{sec:threat-model}

\noindent
\textbf{Attack scenario and Objectives.}
We consider practical scenarios where LLM agents operate in real-world environments and ingest published content. 
The threat arises when an adversarial agent-driven crawler or automated LLM-based pipeline seeks a faithful compressed view of the content for unauthorized downstream reuse.
Accordingly, the adversary's objective is to extract and condense the text flawlessly.
In contrast, the defender's objective is to proactively disrupt this compression process, publishing a protected version of the content that is with minimal human-visible surface change but becomes semantically unreliable once compressed.
This reflects realistic constraints, where defenses must operate stealthily to avoid detection or disruption for legitimate human readers.
Unlike prompt-injection attacks that manipulate immediate model outputs, our defense uniquely targets the fidelity of the agent's content reuse.




\noindent
\textbf{Victim and Defender Capabilities.}
The victim is the content owner whose high-value textual assets face unauthorized scraping and use by agentic crawlers.
The content owner acts proactively as the defender and can only preprocess the text before publication.
They cannot rely on platform-level controls or intervene in the adversary's external agent workflows.
Furthermore, the defender operates under a restricted knowledge setting.
When content is published, the defender cannot foresee which specific agent will later scrape and use the content, nor access the parameters or implementation of any adversary's compression module, since such modules are often closed-source commercial products.
To overcome this target unpredictability, the defender assumes white-box access only to a set of surrogate open-source compressors.
The defender uses these surrogates to optimize invisible perturbations, aiming for robust transferability to unknown black-box targets in the wild.

\noindent
\textbf{Adversary and Capabilities.}
The adversary deploys automated crawlers, such as headless browsers or agent-driven scrapers, to extract online digital assets from hosting platforms and feed them into an agent workflow.
We assume a strong adversary that can freely choose its compression model and downstream use procedures.
By default, it has no prior knowledge that an asset is protected and treats the harvested content as authentic, pristine input.
To preserve information completeness for downstream tasks, the adversary forwards the extracted text to its compression module verbatim, without aggressive preprocessing or sanitization that could discard semantically useful tokens.
Crucially, the adversary's capabilities are restricted to the client side, and they cannot modify the defender's published content at the source.

\subsection{Problem Formulation}
\label{sec:problem-formulation}

To define the protection mechanism, we first formalize the standard agent workflow. 
Given an input content $x$, the agent first produces a compressed representation $c$ via a compression module $f$, and then generates downstream output $y$ through a downstream module $g$ conditioned on $c$:
\begin{equation}
c=f(x;\pi_f), \qquad y=g(q,c,r;\pi_g),
\label{eq:agent-workflow}
\end{equation}
where $\pi_f,\pi_g$ are system instructions, $q$ is the user query, and $r$ is the task context, such as tool states, retrieved information, or conversation metadata.

The defense constructs protected text $x'$ whose compressed representation $c'$ is less faithful while its human-visible surface difference from $x$ remains small.
It constructs an invisible perturbation
\begin{equation}
a=(p,L,z_{1:L}), \qquad z_i\in\mathcal{V}_{\mathrm{inv}},
\label{eq:perturbation-instance}
\end{equation}
where $p$ is the insertion position, $L$ is the perturbation length, $z_{1:L}$ is the invisible-token sequence, and $\mathcal{V}_{\mathrm{inv}}$ is the invisible-token set following prior studies~\citep{boucher2022bad}. Let $I(x,a)$ be the insertion operator and $x'=I(x,a)$ the protected text. Under \autoref{eq:agent-workflow}, $x'$ yields $c'=f(x';\pi_f)$ and downstream output $y'$.
We consider two complementary degradation objectives. Textual degradation $\mathcal{D}_{\mathrm{text}}(c')$ captures malformed, repetitive, truncated, or structurally disrupted compression. Information degradation $\mathcal{D}_{\mathrm{info}}(x,c')$ captures the loss of task-relevant facts, entities, code semantics, or dialogue states. The protection objective is
\[
\begin{aligned}
a^\star
=
\operatorname*{arg\,max}_{a}\quad
&
\lambda_{\mathrm{text}}\mathcal{D}_{\mathrm{text}}(c')
+
\lambda_{\mathrm{info}}\mathcal{D}_{\mathrm{info}}(x,c') \\
\text{s.t.}\quad
&
z_i\in\mathcal{V}_{\mathrm{inv}},\quad
\mathrm{HVID}(x,x')\le \epsilon_{\mathrm{vis}} .
\end{aligned}
\]
HVID is a relative input-side difference score, and its full definition is given in \autoref{app:metric-details}.
The final protected output is $x^\star=I(x,a^\star)$.

\begin{figure*}[!t]
    \centering
    \includegraphics[width=0.95\textwidth]{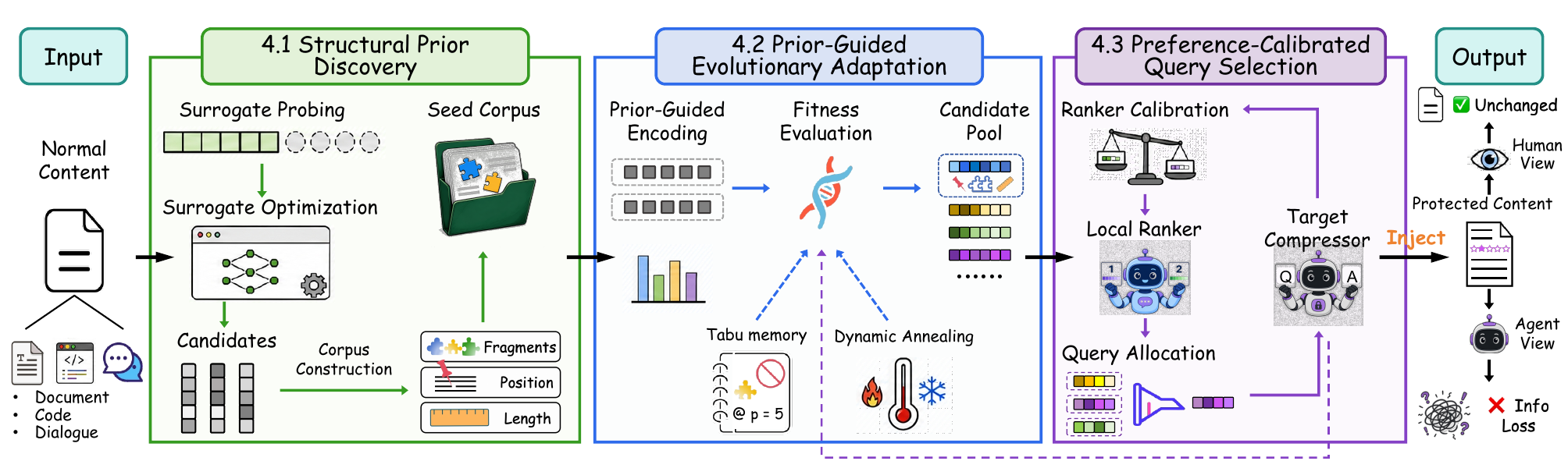}
    \caption{Overview of \sys. \sys discovers structural priors from accessible compressors, adapts them through prior-guided evolution, and allocates closed-source queries via preference-calibrated selection to generate adapted protective perturbations.}
    \label{fig:sys_overview}
    \vspace{-2mm}
\end{figure*}

\section{Methodology}\label{sec:4_Methodology}

Building on the compression degradation objectives defined above, we propose \textbf{\sys} (\textbf{C}ompression-\textbf{A}ware \textbf{P}rotective \textbf{E}volution), an active content protection framework for LLM-based agent workflows. Without changing the human-visible surface form, \sys optimizes invisible perturbations so that the compressed representation of protected content suffers structural degradation and key-information loss, reducing downstream reusability.~\autoref{fig:sys_overview} illustrates the overall workflow.

\sys contains three stages.
\textit{Structural Prior Discovery} uses an accessible surrogate compressor to construct a seed corpus of effective perturbations and extract structural descriptors for later adaptation (\autoref{sec:whitebox-prior}).
\textit{Prior-guided Evolutionary Adaptation} starts from these priors and searches for candidates through structure-aware variation and selection (\autoref{sec:blackbox-evolution}).
\textit{Preference-Calibrated Query Selection} calibrates a local preference ranker with target compressor feedback and allocates limited queries (\autoref{sec:proxy-search}).
During the adaptation process, the latter two stages form a closed loop in which target feedback updates both ranker calibration and evolutionary selection.
Finally, \sys outputs the protected text injected with the invisible perturbation.
The full procedure is given in~\autoref{app:cape-algorithm}.



\subsection{Structural Prior Discovery}
\label{sec:whitebox-prior}


This stage constructs a seed corpus of effective invisible perturbations by exploiting a parameter- and gradient-accessible surrogate compressor.
We first optimize perturbations to disrupt the surrogate's generation distribution, and then extract transferable structural priors from high-scoring candidates rather than relying on a single optimal string.

\noindent
{\bf Surrogate Probing.}
Unlike target-string adversarial attacks that force a predefined output, \sys perturbs the generation distribution away from readable and information-preserving compression.

To construct a stable signal for optimization, we design a teacher-forced probing mechanism.
For each candidate $a$, \sys appends a fixed probe continuation $u_{1:T}$ after $I(x,a)$, where the probe is constructed once from valid non-special surrogate tokenizer IDs and shared across all candidates.
At probe position $t$, the surrogate compressor's next-token distribution over token $v$ is defined as:
\begin{equation}
q_t^{a}(v)=p_\theta(v\mid I(x,a),u_{<t}),
\label{eq:probe-distribution}
\end{equation}
where $u_{<t}$ denotes the preceding probe context.
The probe is used only to evaluate perturbation-induced changes in future generation distributions.

\noindent
{\bf Surrogate Optimization.}
Using the probe signal, \sys constructs a multi-term objective to search for perturbations that can induce severe compression degradation.
Let $\mathcal{V}_{\mathrm{anom}}$ denote an anomalous continuation set (including noisy, rare, control-related tokens, and abnormal-pattern tokens). Let $\mathcal{V}_{\mathrm{lang}}$ denote frequent natural-language continuation tokens that support fluent compression. \sys searches for a perturbation $a$ by maximizing:
\begin{equation}
\begin{aligned}
\operatorname*{arg\,max}_{a}\;
\frac{1}{T}\sum_{t=1}^{T}
\Big[
& H(q_t^{a})
+ \lambda_a \log q_t^{a}(\mathcal{V}_{\mathrm{anom}}) \\
& - \lambda_l \log q_t^{a}(\mathcal{V}_{\mathrm{lang}})
\Big].
\end{aligned}
\label{eq:whitebox-objective}
\end{equation}
where $q_t^{a}(\mathcal{V})=\sum_{v\in\mathcal{V}}q_t^{a}(v)$ is the marginal probability assigned to a token set, and $H(q_t^{a})$ denotes distribution entropy.
The three terms serve complementary roles: entropy increases future generation uncertainty, the anomalous continuation term shifts probability mass toward atypical token regions, and the language suppression term penalizes fluent natural-language continuations. 

\noindent
{\bf Corpus Construction.}
\sys distills the high-scoring candidates into a set of transferable structural priors and constructs the seed corpus.
Specifically, it retains three complementary types of priors.
\ding{182} Inspired by contrastive attribution~\citep{sarti2024quantifying}, we identify \textit{degradation-associated local fragments}, contiguous segments whose presence is consistently linked to higher degradation.
\ding{183} We mine \textit{fragment co-occurrence patterns} to capture fragment pairs whose joint presence amplifies degradation beyond their individual effects.
\ding{184} We record \textit{position-length compatibility} by locating high-response regions in the discrete space of insertion positions and perturbation lengths.

These priors form structural descriptors, summarizing degradation-relevant local fragments, fragment co-occurrences, and position--length attributes for later variation and prior-consistency scoring. For each retained seed, the corpus stores its descriptors and surrogate-side performance scores. Details of this stage are provided in~\autoref{app:whitebox-loss}.

\subsection{Prior-Guided Evolutionary Adaptation}
\label{sec:blackbox-evolution}

This stage adapts the surrogate-derived seed corpus to the target compressor through prior-guided evolutionary search. Because target gradients are unavailable, \sys encodes candidates with structural descriptors, evaluates them using a regularized fitness function, and balances exploration and exploitation through dynamic annealing.




\noindent
{\bf Prior-Guided Encoding.}
Target-side adaptation is a gradient-free discrete search over token sequences, insertion positions, and lengths.
To efficiently navigate this space, we build on genetic search~\citep{cho2024typos} but replace character level mutation with prior-guided evolution.
Each candidate is encoded as $g=(a,\mathbf{m},h)$.
$\mathbf{m}$ stores matched structural descriptors from~\autoref{sec:whitebox-prior}, and $h$ records evaluation history.
Then, recombination and mutation operate over this encoding, with the matched $\mathbf{m}$ steering variation toward prior-supported structures rather than arbitrary characters.
Detailed operators are introduced in~\autoref{app:blackbox-evolution}.


\noindent
{\bf Regularized Fitness Evaluation.}
To reliably evaluate candidates and mitigate noisy feedback from closed-source compressors, we rank them with a fitness score combining observed degradation, prior consistency, stability, and exploration memory:
\[
\widetilde{F}_t(g)
=
D_t(g)+\lambda_p S_p(g)-\lambda_u U_t(g)-\lambda_r R_{\mathrm{tabu}}(g).
\]
where $D_t(g)$ measures target compressor degradation.
The remaining terms regularize the search: $S_p(g)$ rewards consistency with surrogate-derived priors, $U_t(g)$ penalizes unstable feedback, and $R_{\mathrm{tabu}}(g)$ discourages revisiting low-value structural regions.
Together, these terms prevent the search from overfitting to single query anomalies.


\noindent
{\bf Dynamic Annealing Mechanism.}
Even with regularized fitness, greedy selection can prematurely concentrate around local optima.
To actively balance the exploitation of current elites with the exploration of novel structures, \sys adjusts the acceptance temperature according to structural diversity and novelty.
The temperature increases when diversity decreases or an offspring introduces a novel structure, allowing exploratory, potentially lower-scoring candidates to remain in the population.
As the population remains diverse and recent generations improve consistently, the temperature decays and the search shifts toward convergence.
Details of this stage are provided in~\autoref{app:blackbox-evolution}.

\subsection{Preference-Calibrated Query Selection}
\label{sec:proxy-search}

This final stage bridges the large pool of evolved candidates and the strict query budgets of target compressors. It calibrates a local preference ranker for low cost scoring, allocates limited queries with a balanced acquisition function, and updates the ranker with target compressor feedback.



\noindent
{\bf Local Ranker Calibration.}
To establish a low cost evaluator, we formulate ranker calibration as a preference ranking problem, focusing on within-pool ordering and avoiding cross model score scale mismatches.
Let $f_\phi(g)$ denote the degradation potential predicted by the local ranker and $D_{\mathrm{tar}}(g)$ denote the degradation measured from target compressor feedback.
To avoid unreliable near-tie comparisons, we construct preference pairs only when their target-side degradation gap exceeds $\tau$.
Let $\mathcal{P}_t$ be the set of significant preference pairs accumulated by round $t$.
For each pair $(g_i,g_j)\in\mathcal{P}_t$, define the direction of target feedback as $y_{ij}=\mathrm{sign}(D_{\mathrm{tar}}(g_i)-D_{\mathrm{tar}}(g_j))$ and the ranker predicted gap as $\Delta_\phi^{ij}=f_\phi(g_i)-f_\phi(g_j)$.
The ranker is calibrated by minimizing the following loss:
\begin{equation}
\mathcal{L}_{\mathrm{align}}(\phi)
=
\sum_{(g_i,g_j)}
\log\!\left(1+\exp(-y_{ij}\Delta_\phi^{ij})\right).
\label{eq:proxy-align}
\end{equation}
This objective trains the ranker to recover relative candidate quality rather than absolute degradation values, yielding a robust signal for selection.





\noindent
{\bf Query Allocation and Update.}
After calibration, \sys selects target compressor queries using an acquisition score that balances predicted effectiveness, ranker uncertainty, and structural coverage:
\begin{equation}
\alpha_t(g)
=
f_\phi(g)
+
\beta_u \mathcal{U}_\phi(g)
+
\beta_n \mathcal{N}_{\mathrm{str}}(g\mid\mathcal{H}_t),
\label{eq:query-priority}
\end{equation}
The three terms correspond to exploitation, calibration, and exploration: $f_\phi(g)$ scores predicted degradation, $\mathcal{U}_\phi(g)$ targets candidates of high ranker uncertainty, and $\mathcal{N}_{\mathrm{str}}(g\mid\mathcal{H}_t)$ rewards structural novelty relative to historical query set $\mathcal{H}_t$, preventing premature concentration on narrow patterns.



The query allocation process forms a closed loop to continually refine the ranker.
In each evolutionary round, \sys scores the candidate pool, submits the Top-$K$ candidates under $\alpha_t(g)$ to the target compressor, and stores evaluated candidates with their target compressor scores.
These records provide new preference pairs for ranker updates, making this stage the dynamic query-selection layer for expensive target compressor evaluations.

\section{Experiments} \label{sec:5_experiment}

We answer two core research questions (RQs).

\noindent 
\textbf{RQ1 (Effectiveness):} Can \sys effectively disrupt compression results and downstream utility?

\noindent 
\textbf{RQ2 (Ablation Study):} What is the contribution of each design to compression degradation?

\subsection{Experimental Setup} \label{sec:5_experiment_setup}

\noindent \textbf{Datasets.}
We evaluate \sys on three content types: long-form documents (Task Haystack~\citep{xu2024stress}), code snippets (CoRE~\citep{xie2026core}, CAB~\citep{kim2026codeassistbench}), and multi-turn dialogues (BABILong~\citep{kuratov2024babilong}, T1~\citep{chakraborty2026t1}).
Each example is evaluated with and without the invisible perturbation under the same evaluation pipeline.

\noindent \textbf{Compressors and Baselines.}
To comprehensively evaluate \sys in realistic content protection scenarios, we cover mainstream model compressor settings, agent workflow compression settings, and commercial application settings.
Only the compression result (including compressed internal representation) can be accessed in any of the above settings.
For model compressors, we evaluate GPT-4.1 and Gemini 3 Flash.
For agent workflows, we use LangGraph~\citep{langchain2024langgraph} as the representative framework.
For commercial applications, we evaluate GitHub Copilot~\citep{github2021copilot}.
We include \textit{Random Invisible} and \textit{Fixed Zero-width} perturbations, to test whether degradation is caused merely by the presence of invisible characters.
We also report two \sys variants: \textit{Direct Transfer (Direct Trans.)}, which removes the guided evolutionary adaptation, and \textit{Prior-Free Evolution (Prior-Free)}, which removes the prior discovery.
External baselines include \textit{TAP}~\citep{mehrotra2024tree} and \textit{HardCom}~\citep{liu2025compressionattack}.
More details of compressors and baselines are in~\autoref{app:model-workflow-details} and~\autoref{app:baseline-details}.

\begin{table*}[t]
\centering
\scriptsize
\setlength{\tabcolsep}{1.15pt}
\renewcommand{\arraystretch}{1.06}

\definecolor{maintabgray}{RGB}{243,243,243}
\definecolor{maintabpurple}{RGB}{232,231,255}
\definecolor{maintabupred}{RGB}{190,40,40}
\definecolor{maintabdown}{RGB}{40,130,70}

\providecommand{\tbbase}[1]{\ensuremath{#1}}
\providecommand{\tbinc}[2]{\ensuremath{#1_{\textcolor{maintabupred}{\scriptscriptstyle #2}}}}
\providecommand{\tbdec}[2]{\ensuremath{#1_{\textcolor{maintabdown}{\scriptscriptstyle #2}}}}
\providecommand{\tbbestinc}[2]{\ensuremath{\mathbf{#1}_{\textcolor{maintabupred}{\scriptscriptstyle #2}}}}
\providecommand{\tbbestdec}[2]{\ensuremath{\mathbf{#1}_{\textcolor{maintabdown}{\scriptscriptstyle #2}}}}
\providecommand{\tbsecinc}[2]{\ensuremath{\underline{#1}_{\textcolor{maintabupred}{\scriptscriptstyle #2}}}}
\providecommand{\tbsecdec}[2]{\ensuremath{\underline{#1}_{\textcolor{maintabdown}{\scriptscriptstyle #2}}}}
\providecommand{\hvidbasebest}[1]{\ensuremath{\mathbf{#1}}}
\providecommand{\hvidbasesec}[1]{\ensuremath{\underline{#1}}}

\providecommand{\hvidinc}[2]{\ensuremath{#1_{\textcolor{maintabdown}{\scriptscriptstyle #2}}}}
\providecommand{\hviddec}[2]{\ensuremath{#1_{\textcolor{maintabupred}{\scriptscriptstyle #2}}}}
\providecommand{\hvidbestinc}[2]{\ensuremath{\mathbf{#1}_{\textcolor{maintabdown}{\scriptscriptstyle #2}}}}
\providecommand{\hvidbestdec}[2]{\ensuremath{\mathbf{#1}_{\textcolor{maintabupred}{\scriptscriptstyle #2}}}}
\providecommand{\hvidsecinc}[2]{\ensuremath{\underline{#1}_{\textcolor{maintabdown}{\scriptscriptstyle #2}}}}
\providecommand{\hvidsecdec}[2]{\ensuremath{\underline{#1}_{\textcolor{maintabupred}{\scriptscriptstyle #2}}}}

\resizebox{\textwidth}{!}{
\begin{tabular}{lcccccccccccc}
\toprule
\multirow{2}{*}{\textbf{Method}}
& \multicolumn{4}{c}{\textbf{Text}}
& \multicolumn{4}{c}{\textbf{Code}}
& \multicolumn{4}{c}{\textbf{Dialogue}} \\
\cmidrule(lr){2-5}\cmidrule(lr){6-9}\cmidrule(lr){10-13}
& \textbf{TD} $\uparrow$ & \textbf{ID} $\uparrow$ & \textbf{OSD} $\uparrow$ & \textbf{HVID} $\downarrow$
& \textbf{TD} $\uparrow$ & \textbf{ID} $\uparrow$ & \textbf{OSD} $\uparrow$ & \textbf{HVID} $\downarrow$
& \textbf{TD} $\uparrow$ & \textbf{ID} $\uparrow$ & \textbf{OSD} $\uparrow$ & \textbf{HVID} $\downarrow$ \\
\midrule

\multicolumn{13}{c}{\textbf{GPT-4.1}} \\
\midrule
Random
& \tbbase{0.5} & \tbbase{1.4} & \tbbase{1.7} & \hvidbasebest{1.4}
& \tbbase{0.2} & \tbbase{0.8} & \tbbase{1.1} & \hvidbasesec{1.1}
& \tbbase{1.0} & \tbbase{1.3} & \tbbase{2.1} & \hvidbasesec{0.9} \\

\rowcolor{maintabgray}
Fixed
& \tbinc{1.5}{1.0} & \tbdec{0.9}{0.5} & \tbinc{1.9}{0.2} & \hvidsecinc{1.9}{0.5}
& \tbinc{1.2}{1.0} & \tbinc{1.4}{0.6} & \tbinc{2.1}{1.0} & \hvidbestdec{0.4}{0.7}
& \tbdec{0.9}{0.1} & \tbdec{0.8}{0.5} & \tbdec{1.4}{0.7} & \hvidbestdec{0.6}{0.3} \\

Direct Trans.
& \tbinc{24.6}{24.1} & \tbinc{30.7}{29.3} & \tbinc{33.9}{32.2} & \hvidinc{6.3}{4.9}
& \tbinc{16.3}{16.1} & \tbinc{19.4}{18.6} & \tbinc{20.9}{19.8} & \hvidinc{4.9}{3.8}
& \tbsecinc{21.4}{20.4} & \tbinc{28.2}{26.9} & \tbinc{25.1}{23.0} & \hvidinc{1.2}{0.3} \\

\rowcolor{maintabgray}
Prior-Free
& \tbsecinc{32.5}{32.0} & \tbinc{39.4}{38.0} & \tbinc{42.1}{40.4} & \hvidinc{5.9}{4.5}
& \tbsecinc{28.4}{28.2} & \tbsecinc{26.8}{26.0} & \tbinc{29.3}{28.2} & \hvidinc{1.8}{0.7}
& \tbinc{20.3}{19.3} & \tbinc{31.6}{30.3} & \tbinc{37.9}{35.8} & \hvidinc{2.9}{2.0} \\

TAP
& \tbinc{14.0}{13.5} & \tbinc{40.2}{38.8} & \tbinc{38.1}{36.4} & \hvidinc{51.6}{50.2}
& \tbinc{13.7}{13.5} & \tbinc{24.1}{23.3} & \tbinc{25.7}{24.6} & \hvidinc{44.9}{43.8}
& \tbinc{9.9}{8.9} & \tbsecinc{34.0}{32.7} & \tbinc{34.2}{32.1} & \hvidinc{50.1}{49.2} \\

\rowcolor{maintabgray}
REGTEXT
& \tbinc{11.8}{11.3} & \tbinc{29.0}{27.6} & \tbinc{26.4}{24.7} & \hvidinc{31.9}{30.5}
& \tbinc{11.2}{11.0} & \tbinc{18.4}{17.6} & \tbinc{21.0}{19.9} & \hvidinc{38.7}{37.6}
& \tbinc{14.3}{13.3} & \tbinc{25.6}{24.3} & \tbinc{24.1}{22.0} & \hvidinc{39.0}{38.1} \\

HardCom
& \tbinc{17.4}{16.9} & \tbsecinc{41.8}{40.4} & \tbsecinc{47.0}{45.3} & \hvidinc{35.2}{33.8}
& \tbinc{21.8}{21.6} & \tbinc{23.1}{22.3} & \tbsecinc{31.8}{30.7} & \hvidinc{26.4}{25.3}
& \tbinc{19.4}{18.4} & \tbinc{33.7}{32.4} & \tbsecinc{42.2}{40.1} & \hvidinc{38.7}{37.8} \\

\rowcolor{maintabpurple}
\textbf{CAPE}
& \tbbestinc{49.2}{48.7} & \tbbestinc{56.4}{55.0} & \tbbestinc{61.9}{60.2} & \hvidinc{3.2}{1.8}
& \tbbestinc{42.8}{42.6} & \tbbestinc{40.6}{39.8} & \tbbestinc{45.9}{44.8} & \hvidinc{1.4}{0.3}
& \tbbestinc{49.3}{48.3} & \tbbestinc{53.5}{52.2} & \tbbestinc{57.6}{55.5} & \hvidinc{3.2}{2.3} \\

\midrule

\multicolumn{13}{c}{\textbf{Gemini 3 Flash}} \\
\midrule
Random
& \tbbase{0.7} & \tbbase{0.9} & \tbbase{1.3} & \hvidbasebest{1.4}
& \tbbase{0.4} & \tbbase{0.6} & \tbbase{0.9} & \hvidbasesec{1.1}
& \tbbase{1.4} & \tbbase{0.8} & \tbbase{2.6} & \hvidbasesec{0.9} \\

\rowcolor{maintabgray}
Fixed
& \tbinc{1.2}{0.5} & \tbdec{0.7}{0.2} & \tbinc{1.8}{0.5} & \hvidsecinc{1.9}{0.5}
& \tbinc{0.6}{0.2} & \tbinc{1.1}{0.5} & \tbinc{1.4}{0.5} & \hvidbestdec{0.4}{0.7}
& \tbinc{1.6}{0.2} & \tbdec{0.6}{0.2} & \tbdec{1.9}{0.7} & \hvidbestdec{0.6}{0.3} \\

Direct Trans.
& \tbinc{18.1}{17.4} & \tbinc{25.3}{24.4} & \tbinc{38.6}{37.3} & \hvidinc{8.2}{6.8}
& \tbinc{9.2}{8.8} & \tbinc{15.1}{14.5} & \tbinc{14.3}{13.4} & \hvidinc{5.4}{4.3}
& \tbinc{22.3}{20.9} & \tbinc{20.8}{20.0} & \tbinc{31.0}{28.4} & \hvidinc{2.8}{1.9} \\

\rowcolor{maintabgray}
Prior-Free
& \tbsecinc{30.4}{29.7} & \tbinc{30.1}{29.2} & \tbinc{41.2}{39.9} & \hvidinc{4.3}{2.9}
& \tbinc{21.4}{21.0} & \tbsecinc{27.3}{26.7} & \tbinc{26.8}{25.9} & \hvidinc{1.8}{0.7}
& \tbsecinc{25.1}{23.7} & \tbinc{27.3}{26.5} & \tbinc{34.1}{31.5} & \hvidinc{2.9}{2.0} \\

TAP
& \tbinc{27.6}{26.9} & \tbinc{33.4}{32.5} & \tbinc{36.8}{35.5} & \hvidinc{49.1}{47.7}
& \tbsecinc{26.5}{26.1} & \tbinc{19.0}{18.4} & \tbinc{23.1}{22.2} & \hvidinc{39.8}{38.7}
& \tbinc{20.9}{19.5} & \tbinc{30.2}{29.4} & \tbsecinc{41.2}{38.6} & \hvidinc{40.3}{39.4} \\

\rowcolor{maintabgray}
REGTEXT
& \tbinc{10.6}{9.9} & \tbinc{24.7}{23.8} & \tbinc{27.6}{26.3} & \hvidinc{33.9}{32.5}
& \tbinc{8.5}{8.1} & \tbinc{15.8}{15.2} & \tbinc{18.1}{17.2} & \hvidinc{26.4}{25.3}
& \tbinc{10.9}{9.5} & \tbinc{25.4}{24.6} & \tbinc{28.3}{25.7} & \hvidinc{35.3}{34.4} \\

HardCom
& \tbinc{13.2}{12.5} & \tbsecinc{36.2}{35.3} & \tbsecinc{43.9}{42.6} & \hvidinc{42.8}{41.4}
& \tbinc{18.4}{18.0} & \tbinc{25.9}{25.3} & \tbsecinc{29.1}{28.2} & \hvidinc{31.2}{30.1}
& \tbinc{18.2}{16.8} & \tbsecinc{33.1}{32.3} & \tbinc{37.2}{34.6} & \hvidinc{42.6}{41.7} \\

\rowcolor{maintabpurple}
\textbf{CAPE}
& \tbbestinc{45.1}{44.4} & \tbbestinc{51.9}{51.0} & \tbbestinc{57.2}{55.9} & \hvidinc{3.4}{2.0}
& \tbbestinc{30.2}{29.8} & \tbbestinc{34.1}{33.5} & \tbbestinc{40.9}{40.0} & \hvidinc{2.6}{1.5}
& \tbbestinc{38.3}{36.9} & \tbbestinc{44.1}{43.3} & \tbbestinc{49.3}{46.7} & \hvidinc{2.4}{1.5} \\

\bottomrule
\end{tabular}
}

\caption{
Main results on closed-source compressors.\scriptsize{(
The symbols $\uparrow$ and $\downarrow$ separately indicate whether a higher or lower value of a specific metric is preferable.
Colored subscripts report absolute changes from `Random', where \textcolor[HTML]{FF4747}{red} indicates a favorable change while \textcolor[HTML]{00823B}{green} indicates an unfavorable change.)}
}
\label{tab:main-results}
\end{table*}

\noindent \textbf{Metrics.}
We report three groups of metrics. 
\ding{182} \textit{Compression-level protection}: Textual Degradation (TD) is the primary metric and measures corruption and reliability loss in compressed outputs; Information Degradation (ID) measures loss of recoverable key information; Output Semantic Drift (OSD) measures semantic drift between reference and protected outputs. 
\ding{183} \textit{Human-visible input difference}: HVID measures the human-visible difference between the original input and the input injected with the invisible perturbation. 
\ding{184} \textit{Workflow-level utility loss}: DRAD measures downstream accuracy drop in LangGraph, while CSD and MOR measure code-level failure and malformed generations in Copilot.
Formal definitions are provided in~\autoref{app:metric-details}.

\noindent \textbf{Implementation.}
Unless otherwise specified, the invisible perturbation length is \(1/20\) of the input length, and target side adaptation uses at most 100 target compressor queries per example. Coefficients are selected by grid search on a held-out validation split and fixed across all experiments: \(\lambda_a=0.75\), \(\lambda_l=0.30\), \(\lambda_p=0.30\), \(\lambda_u=0.15\), \(\lambda_r=0.10\), \(\beta_u=0.25\), and \(\beta_n=0.15\).
More details for these default values are in~\autoref{app:implementation-details}.

\subsection{RQ1: Effectiveness of \sys} \label{sec:rq1}


\noindent
\textbf{Experimental Design.}
Real-world agent usage is heavily concentrated and dominated by a small number of mainstream pipelines, such as GitHub Copilot~\cite{stackoverflow2025ai}, so that pre-profiling a few representative targets and generating perturbations can cover a large fraction of the realistic deployment risks.
In this section, we evaluate \sys across four target compressors and three content-type dataset groups that together span this mainstream agent ecosystem.
We test GPT-4.1 and Gemini 3 Flash on all datasets.
For LangGraph, we use context-related Task Haystack and T1 tasks covering factual question answering, state tracking, and constraint recovery, reflecting agent workflows over compressed long-document and dialogue-history representations.
For GitHub Copilot, we evaluate code completion and bug fixing on CoRE and CAB, following common usage scenarios of this commercial assistant.




\noindent
\textbf{Analysis on Compression Models.}
As shown in~\autoref{tab:main-results}, \sys outperforms all baselines on both GPT-4.1 and Gemini 3 Flash.


The GPT-4.1 results show that effective protection requires both transferable priors and target side adaptation.
\textit{Direct Transfer} outperforms sanity checks, suggesting that the structural prior discovery stage extracts partially transferable degradation patterns. However, its gap from \sys indicates that these patterns cannot serve as universal perturbations under architectural mismatch between surrogate and target compressors.
\textit{Prior-Free Evolution} and \textit{TAP} remain limited, showing that feedback-driven search is inefficient without compression-specific structure.
By combining prior-guided evolution with target feedback, \sys improves information loss by up to 62.6\% over the best baseline, indicating that target side adaptation is necessary beyond seed reuse.

The Gemini 3 Flash results further confirm the generality of \sys.
Compared with the strongest external baseline, \sys improves TD by 241.7\% and semantic drift on long-form text by 30.3\%.
More importantly, \sys keeps HVID in the low single digits, unlike external baselines with much larger input differences, showing that it disrupts compressed outputs while introducing almost no human-visible change to the protected content.

Across content types, long-form text is most affected, dialogue is intermediate, and code is more resistant.
This reflects code's stricter structural constraints, where syntax, identifiers, and local dependencies make compression less flexible.
Nevertheless, \sys remains strongest on code with 42.8\% TD, showing that its protection extends to structure-sensitive code content.

\begin{table}[t]
\centering
\scriptsize
\setlength{\tabcolsep}{8pt}
\renewcommand{\arraystretch}{0.98}

\definecolor{capetabgray}{RGB}{243,243,243}
\definecolor{capetabpurple}{RGB}{232,231,255}
\definecolor{capetabupred}{RGB}{190,40,40}
\definecolor{capetabdown}{RGB}{40,130,70}

\providecommand{\lgbase}[1]{\ensuremath{#1}}
\providecommand{\lgup}[2]{\ensuremath{#1_{\textcolor{capetabupred}{\scriptscriptstyle \uparrow #2}}}}
\providecommand{\lgdown}[2]{\ensuremath{#1_{\textcolor{capetabdown}{\scriptscriptstyle \downarrow #2}}}}
\providecommand{\lgbestup}[2]{\ensuremath{\mathbf{#1}_{\textcolor{capetabupred}{\scriptscriptstyle \uparrow #2}}}}
\providecommand{\lgsecup}[2]{\ensuremath{\underline{#1}_{\textcolor{capetabupred}{\scriptscriptstyle \uparrow #2}}}}
\providecommand{\lgsecdw}[2]{\ensuremath{\underline{#1}_{\textcolor{capetabdown}{\scriptscriptstyle \downarrow #2}}}}

\begin{tabular}{lccc}
\toprule
\textbf{Method}
& \textbf{TD} $\uparrow$
& \textbf{DRAD} $\uparrow$
& \textbf{HVID} $\downarrow$ \\
\midrule

Random
& \lgbase{0.6}
& \lgbase{2.1}
& \textbf{\lgbase{1.1}} \\

\rowcolor{capetabgray}
HardCom
& \lgsecup{20.7}{20.1}
& \lgsecup{26.2}{24.1}
& \lgdown{46.1}{45.0} \\

\rowcolor{capetabpurple}
\textbf{CAPE}
& \lgbestup{52.9}{52.3}
& \lgbestup{59.7}{57.6}
& \lgsecdw{2.8}{1.7} \\

\bottomrule
\end{tabular}

\caption{
LangGraph workflow evaluation.
}
\label{tab:langgraph-agent}
\end{table}

\noindent
\textbf{Disruption in Agent Workflows.}
\autoref{tab:langgraph-agent} shows that \sys remains highly effective inside an agent workflow.
Compared with \textit{HardCom}, \sys achieves higher TD and drastically lower HVID, delivering stronger protection with orders-of-magnitude better human-visible preservation. Its DRAD of up to 59.7\% confirms severe degradation of facts, states, and constraints essential for downstream tasks, effectively limiting the reuse of compressed high-value content.

\begin{table}[t]
\centering
\scriptsize
\setlength{\tabcolsep}{4pt}
\renewcommand{\arraystretch}{0.98}

\definecolor{copitabgray}{RGB}{243,243,243}
\definecolor{copitabpurple}{RGB}{232,231,255}
\definecolor{copitabred}{RGB}{190,40,40}
\definecolor{copitabdown}{RGB}{40,130,70}

\providecommand{\copb}[1]{\ensuremath{#1}}
\providecommand{\copup}[2]{\ensuremath{#1_{\textcolor{copitabred}{\scriptscriptstyle \uparrow #2}}}}
\providecommand{\copdown}[2]{\ensuremath{#1_{\textcolor{copitabdown}{\scriptscriptstyle \downarrow #2}}}}
\providecommand{\cops}[2]{\ensuremath{\underline{#1}_{\textcolor{copitabred}{\scriptscriptstyle \uparrow #2}}}}
\providecommand{\copbestv}[2]{\ensuremath{\mathbf{#1}_{\textcolor{copitabred}{\scriptscriptstyle \uparrow #2}}}}
\providecommand{\copsecdw}[2]{\ensuremath{\underline{#1}_{\textcolor{copitabdown}{\scriptscriptstyle \downarrow #2}}}}

\begin{tabular}{@{}lcccc@{}}
\toprule
\textbf{Method}
& \textbf{TD} $\uparrow$
& \textbf{CSD} $\uparrow$
& \textbf{MOR} $\uparrow$
& \textbf{HVID} $\downarrow$ \\
\midrule

Random
& \copb{0.3}
& \copb{1.2}
& \copb{0.8}
& \textbf{\copb{1.6}} \\

\rowcolor{copitabgray}
HardCom
& \cops{10.4}{10.1}
& \cops{4.5}{3.3}
& \cops{2.1}{1.3}
& \copdown{42.7}{41.1} \\

\rowcolor{copitabpurple}
\textbf{CAPE}
& \copbestv{32.9}{32.6}
& \copbestv{16.4}{15.2}
& \copbestv{38.5}{37.7}
& \copsecdw{3.4}{1.8} \\

\bottomrule
\end{tabular}

\caption{
Results on GitHub Copilot.
}
\label{tab:copilot-eval}
\end{table}


Similarly,~\autoref{tab:copilot-eval} shows that \sys remains effective against a commercial assistant with inaccessible model internals. 
This shows \sys effectively weakens Copilot's recovery of executable intent and structural constraints without visible original code context rewriting, confirming seamless transferability from controlled compression settings to practical commercial coding assistants.

\begin{figure}[t]
    \centering
    \includegraphics[width=0.98\columnwidth]{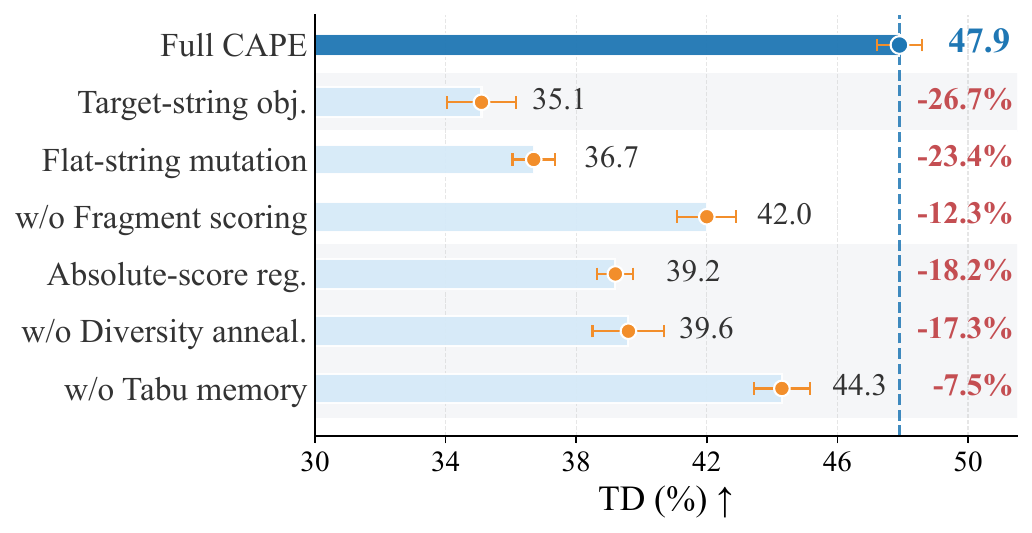}
    \caption{
    Component ablation results.
    }
    \label{fig:ablation}
\end{figure}

Since the specific downstream compressor remains unknown to the defender at publication time (\autoref{sec:threat-model}), \autoref{app:compositional-transfer} further shows that perturbations independently designed for different targets can be composed into a single perturbation that simultaneously induces significant degradation across all four targets, illustrating the effectiveness of \sys in facing several mainstream agent compressors in deployment scenarios.

\subsection{RQ2: Ablation Study} \label{sec:rq2}

\noindent\textbf{Experimental Design.}
We conduct a systematic ablation study by keeping the \sys pipeline intact while replacing or removing key components.
We evaluate six variants: \textit{Target-string objective}, \textit{Flat-string mutation}, \textit{w/o Fragment scoring}, \textit{Absolute-score regression}, \textit{w/o Diversity annealing}, and \textit{w/o Tabu memory}.
These variants examine three core mechanisms of \sys (i.e., distributional surrogate optimization, prior-guided variation and preference calibrated ranker).
\autoref{fig:ablation} reports TD of each variant, with red marking each variant's degradation relative to the complete \sys.
Variant details and stage-level ablation results are in~\autoref{app:additional-experiments}.



\noindent
\textbf{Result Analysis.}
From~\autoref{fig:ablation}, replacing or removing any key mechanism of \sys leads to significant performance degradation.
Specifically, replacing the distributional surrogate objective with a conventional target string reduces TD by 26.7\%, showing that fixed-output matching fails to learn transferable degradation conditions.
For perturbation representation, \textit{Flat-string mutation} and \textit{w/o Fragment scoring} also cause large drops, with the latter reducing TD by 23.4\%, indicating that effective invisible perturbations require structured candidate modeling rather than ordinary string editing.
For feedback and search control, \textit{Absolute-score reg.}, \textit{w/o Diversity annealing}, and \textit{w/o Tabu memory} all underperform, confirming the need for preference-calibrated and diversity-aware target-side search.
Together, these results support the three-stage design of \sys.

\section{Conclusion} \label{sec:6_conclusion}


In this paper, we present \sys, a framework that protects high-value online content from unauthorized agentic scraping without changing its human-visible surface form.
\sys identifies context compression as an overlooked defense layer in agent pipelines and injects invisible perturbations into source content to induce severe semantic loss after compression.
Experiments across three content types and four target settings show that \sys improves information loss by up to 75.8\% over the strongest baseline while keeping protected content visually indistinguishable from the original.
These results instantiate context compression as a practical defense layer and motivate future research on content protection in the agent era.

\newpage

\section*{Limitations}

This work aims to reveal and instantiate context compression as a defense layer for online content in the agent era, rather than to deliver a deployment-ready product.
We discuss two main limitations of this work.

\noindent
\textbf{Compression and Agent Coverage.}
Our evaluation spans four representative compression methods including the compression modules of two real-world agent pipelines (i.e., LangGraph and GitHub Copilot).
The agent ecosystem evolves rapidly.
Although \sys is designed to be compressor-independent, the observations of this paper may not hold on some emerging compression methods, given that future compression methods may respond to disturbances differently than the compressor studied in this paper.
To mitigate this risk, we publicly release the prototype of \sys and necessary experimental results to facilitate replication and recalibration on new compression methods in the future.

\noindent
\textbf{Adaptive Adversaries.}
A determined adversary could try to strip zero-width characters before compression, for instance, through Unicode normalization or whitespace sanitization.
Furthermore, as agentic crawlers, applying such preprocessing by default could corrupt legitimate content with formatting-sensitive tokens (e.g., code and structured markup).
This paper aims to reveal context compression as a novel content protection means.
We treat counter-stripping defenses (including redundant and semantics-bound perturbations) as a potential extension direction and leave a systematic adaptive-attacker study to future work.

\section*{Ethical Considerations}

This work studies how content owners can defend their material against unauthorized scraping by agents, and we do not foresee direct ethical harm from the technique itself.
All experiments use publicly available content sources and open or commercial compression services.
The study involves no personal or sensitive information and raises no IRB-level concerns.
We acknowledge that any active protection mechanism could be misused, for example, to obstruct legitimate auditing, accessibility services, or scholarly crawlers.
\sys is positioned as a defense method from the perspective of content owners rather than as an offensive tool, and its perturbations stay within invisible Unicode regions that leave the human-visible text unchanged.
We encourage platforms and standards bodies to develop transparent disclosure mechanisms so that compression-layer protections remain compatible with responsible data use, scholarly inspection, and downstream interoperability.

\bibliography{custom}

\appendix
\newpage
\section{Appendices}

\subsection{Appendix Overview}
\label{app:overview}

The appendix is organized as follows:
\begin{itemize}[leftmargin=*]
    \item \autoref{app:additional-experiments} reports additional experiments and analyses, including compositional transfer, robustness across compression methods, long-context stress tests, query efficiency, perturbation-length sensitivity, and position-length priors.
    \item \autoref{app:cape-algorithm} summarizes the overall procedure of \sys.
    \item \autoref{app:whitebox-loss}, \autoref{app:blackbox-evolution}, and \autoref{app:proxy-search} provide details of Structural Prior Discovery, Prior-Guided Evolutionary Adaptation, and Preference-Calibrated Query Selection, respectively.
    \item \autoref{app:implementation-details} describes implementation settings and hyperparameter grid search.
    \item \autoref{app:dataset-details} and \autoref{app:model-workflow-details} provide dataset, model, and workflow details. \autoref{app:metric-details} defines all evaluation metrics. \autoref{app:baseline-details} describes baseline adaptations and fairness controls.
    \item \autoref{app:discussion} discusses broader implications for content protection in the agent era.
\end{itemize}

\subsection{Additional Experiments and Analyses}
\label{app:additional-experiments}

This section provides additional experiments that complement the main results in \autoref{sec:5_experiment}. 
These analyses further examine the robustness, query efficiency, and structural mechanisms of \sys across compression methods, target compressors, content types, and perturbation configurations.

\noindent \textbf{Surrogate-Access Compressor Results}
\label{app:surrogate-access-results}

Although the main experiments focus on query-only target compressors and practical agent workflows, we additionally evaluate \sys in surrogate-access settings.
This evaluation serves two purposes.
First, several existing baselines are designed for parameter-accessible compressors, and therefore require a separate comparison under the same access condition.
Second, this setting tests whether \textit{Structural Prior Discovery}, the first stage of \sys, is itself an effective protection mechanism and can produce high-quality priors for later target-side adaptation.

We use Llama3-8B and Qwen3-8B as parameter-accessible compressors.
Because the target compressor is directly accessible in this setting, the target compressor itself serves as the surrogate compressor.
Accordingly, \sys reduces to its first stage and requires no target-side evolutionary adaptation or preference-calibrated query selection.
We denote this stage-1-only setting as \sys-WPL, which retains only \textit{Structural Prior Discovery}.
The corresponding baselines include \textit{I-GCG}~\citep{jia2024improved}, \textit{REGTEXT}~\citep{java2025towards}, and \textit{SoftCom}~\citep{liu2025compressionattack}, which are all adapted to the parameter-accessible compressor setting.
We also include two invisible-character controls, \textit{Random Invisible} and \textit{Fixed Zero-width}, to test whether degradation is caused merely by the presence of invisible characters.
All methods follow the same paired evaluation protocol, perturbation budget, decoding configuration, and human-visible surface-difference measurement as the main experiments.

\begin{table*}[t]
\centering
\scriptsize
\setlength{\tabcolsep}{1.15pt}
\renewcommand{\arraystretch}{1.06}

\definecolor{maintabgray}{RGB}{243,243,243}
\definecolor{maintabpurple}{RGB}{232,231,255}
\definecolor{maintabupred}{RGB}{190,40,40}
\definecolor{maintabdown}{RGB}{40,130,70}

\providecommand{\tbbase}[1]{\ensuremath{#1}}
\providecommand{\tbinc}[2]{\ensuremath{#1_{\textcolor{maintabupred}{\scriptscriptstyle #2}}}}
\providecommand{\tbdec}[2]{\ensuremath{#1_{\textcolor{maintabdown}{\scriptscriptstyle #2}}}}
\providecommand{\tbbestinc}[2]{\ensuremath{\mathbf{#1}_{\textcolor{maintabupred}{\scriptscriptstyle #2}}}}
\providecommand{\tbbestdec}[2]{\ensuremath{\mathbf{#1}_{\textcolor{maintabdown}{\scriptscriptstyle #2}}}}
\providecommand{\tbsecinc}[2]{\ensuremath{\underline{#1}_{\textcolor{maintabupred}{\scriptscriptstyle #2}}}}
\providecommand{\tbsecdec}[2]{\ensuremath{\underline{#1}_{\textcolor{maintabdown}{\scriptscriptstyle #2}}}}
\providecommand{\hvidbasebest}[1]{\ensuremath{\mathbf{#1}}}
\providecommand{\hvidbasesec}[1]{\ensuremath{\underline{#1}}}

\providecommand{\hvidinc}[2]{\ensuremath{#1_{\textcolor{maintabdown}{\scriptscriptstyle #2}}}}
\providecommand{\hviddec}[2]{\ensuremath{#1_{\textcolor{maintabupred}{\scriptscriptstyle #2}}}}
\providecommand{\hvidbestinc}[2]{\ensuremath{\mathbf{#1}_{\textcolor{maintabdown}{\scriptscriptstyle #2}}}}
\providecommand{\hvidbestdec}[2]{\ensuremath{\mathbf{#1}_{\textcolor{maintabupred}{\scriptscriptstyle #2}}}}
\providecommand{\hvidsecinc}[2]{\ensuremath{\underline{#1}_{\textcolor{maintabdown}{\scriptscriptstyle #2}}}}
\providecommand{\hvidsecdec}[2]{\ensuremath{\underline{#1}_{\textcolor{maintabupred}{\scriptscriptstyle #2}}}}

\resizebox{\textwidth}{!}{
\begin{tabular}{lcccccccccccc}
\toprule
\multirow{2}{*}{\textbf{Method}}
& \multicolumn{4}{c}{\textbf{Text}}
& \multicolumn{4}{c}{\textbf{Code}}
& \multicolumn{4}{c}{\textbf{Dialogue}} \\
\cmidrule(lr){2-5}\cmidrule(lr){6-9}\cmidrule(lr){10-13}
& \textbf{TD} $\uparrow$ & \textbf{ID} $\uparrow$ & \textbf{OSD} $\uparrow$ & \textbf{HVID} $\downarrow$
& \textbf{TD} $\uparrow$ & \textbf{ID} $\uparrow$ & \textbf{OSD} $\uparrow$ & \textbf{HVID} $\downarrow$
& \textbf{TD} $\uparrow$ & \textbf{ID} $\uparrow$ & \textbf{OSD} $\uparrow$ & \textbf{HVID} $\downarrow$ \\
\midrule

\multicolumn{13}{c}{\textbf{Llama3-8B}} \\
\midrule
Random
& \tbbase{1.9} & \tbbase{1.5} & \tbbase{2.1} & \hvidbasebest{1.4}
& \tbbase{2.2} & \tbbase{1.1} & \tbbase{1.7} & \hvidbasesec{1.1}
& \tbbase{2.4} & \tbbase{1.3} & \tbbase{1.9} & \hvidbasesec{0.9} \\

\rowcolor{maintabgray}
Fixed
& \tbinc{2.4}{0.5} & \tbinc{2.1}{0.6} & \tbinc{2.5}{0.4} & \hvidsecinc{1.9}{0.5}
& \tbinc{2.5}{0.3} & \tbinc{1.4}{0.3} & \tbinc{1.9}{0.2} & \hvidbestdec{0.4}{0.7}
& \tbinc{2.9}{0.5} & \tbinc{2.4}{1.1} & \tbinc{3.7}{1.8} & \hvidbestdec{0.6}{0.3} \\

I-GCG
& \tbsecinc{19.2}{17.3} & \tbinc{51.6}{50.1} & \tbinc{58.6}{56.5} & \hvidinc{48.7}{47.3}
& \tbinc{27.4}{25.2} & \tbinc{34.3}{33.2} & \tbinc{39.2}{37.5} & \hvidinc{37.0}{35.9}
& \tbsecinc{30.1}{27.7} & \tbinc{38.9}{37.6} & \tbinc{46.3}{44.4} & \hvidinc{42.0}{41.1} \\

\rowcolor{maintabgray}
REGTEXT
& \tbinc{11.3}{9.4} & \tbinc{42.2}{40.7} & \tbinc{36.8}{34.7} & \hvidinc{37.3}{35.9}
& \tbsecinc{30.4}{28.2} & \tbinc{39.0}{37.9} & \tbinc{48.1}{46.4} & \hvidinc{28.4}{27.3}
& \tbinc{21.6}{19.2} & \tbinc{41.1}{39.8} & \tbinc{42.3}{40.4} & \hvidinc{32.3}{31.4} \\

SoftCom
& \tbinc{14.6}{12.7} & \tbsecinc{73.7}{72.2} & \tbsecinc{70.9}{68.8} & \hvidinc{39.4}{38.0}
& \tbinc{18.2}{16.0} & \tbsecinc{50.8}{49.7} & \tbsecinc{59.6}{57.9} & \hvidinc{24.9}{23.8}
& \tbinc{22.8}{20.4} & \tbsecinc{68.6}{67.3} & \tbsecinc{75.1}{73.2} & \hvidinc{28.4}{27.5} \\

\rowcolor{maintabpurple}
\textbf{CAPE-WPL}
& \tbbestinc{62.3}{60.4} & \tbbestinc{81.7}{80.2} & \tbbestinc{89.5}{87.4} & \hvidinc{6.7}{5.3}
& \tbbestinc{61.4}{59.2} & \tbbestinc{65.0}{63.9} & \tbbestinc{72.3}{70.6} & \hvidinc{4.2}{3.1}
& \tbbestinc{67.9}{65.5} & \tbbestinc{76.7}{75.4} & \tbbestinc{82.9}{81.0} & \hvidinc{7.1}{6.2} \\

\midrule

\multicolumn{13}{c}{\textbf{Qwen3-8B}} \\
\midrule
Random
& \tbbase{0.8} & \tbbase{1.9} & \tbbase{1.6} & \hvidbasebest{1.4}
& \tbbase{1.4} & \tbbase{1.1} & \tbbase{0.9} & \hvidbasesec{1.1}
& \tbbase{1.9} & \tbbase{1.2} & \tbbase{2.0} & \hvidbasesec{0.9} \\

\rowcolor{maintabgray}
Fixed
& \tbinc{2.7}{1.9} & \tbdec{1.5}{0.4} & \tbinc{2.9}{1.3} & \hvidsecinc{1.9}{0.5}
& \tbinc{3.5}{2.1} & \tbdec{0.9}{0.2} & \tbinc{1.8}{0.9} & \hvidbestdec{0.4}{0.7}
& \tbinc{2.3}{0.4} & \tbinc{1.2}{0.0} & \tbinc{2.9}{0.9} & \hvidbestdec{0.6}{0.3} \\

I-GCG
& \tbsecinc{28.2}{27.4} & \tbinc{40.3}{38.4} & \tbinc{41.5}{39.9} & \hvidinc{37.3}{35.9}
& \tbsecinc{26.0}{24.6} & \tbinc{29.7}{28.6} & \tbinc{37.8}{36.9} & \hvidinc{30.6}{29.5}
& \tbsecinc{30.8}{28.9} & \tbinc{34.7}{33.5} & \tbinc{49.1}{47.1} & \hvidinc{42.9}{42.0} \\

\rowcolor{maintabgray}
REGTEXT
& \tbinc{21.1}{20.3} & \tbinc{33.5}{31.6} & \tbinc{35.2}{33.6} & \hvidinc{36.8}{35.4}
& \tbinc{15.2}{13.8} & \tbinc{21.6}{20.5} & \tbinc{27.3}{26.4} & \hvidinc{28.3}{27.2}
& \tbinc{28.4}{26.5} & \tbinc{39.3}{38.1} & \tbinc{35.5}{33.5} & \hvidinc{32.4}{31.5} \\

SoftCom
& \tbinc{20.3}{19.5} & \tbsecinc{67.1}{65.2} & \tbsecinc{64.7}{63.1} & \hvidinc{32.4}{31.0}
& \tbinc{17.5}{16.1} & \tbsecinc{48.9}{47.8} & \tbsecinc{53.1}{52.2} & \hvidinc{28.7}{27.6}
& \tbinc{22.1}{20.2} & \tbsecinc{62.7}{61.5} & \tbsecinc{70.8}{68.8} & \hvidinc{29.6}{28.7} \\

\rowcolor{maintabpurple}
\textbf{CAPE-WPL}
& \tbbestinc{60.7}{59.9} & \tbbestinc{78.8}{76.9} & \tbbestinc{84.2}{82.6} & \hvidinc{4.6}{3.2}
& \tbbestinc{59.1}{57.7} & \tbbestinc{62.7}{61.6} & \tbbestinc{70.3}{69.4} & \hvidinc{3.1}{2.0}
& \tbbestinc{65.8}{63.9} & \tbbestinc{70.1}{68.9} & \tbbestinc{87.0}{85.0} & \hvidinc{2.8}{1.9} \\

\bottomrule
\end{tabular}
}

\caption{
Appendix results on open-source models.
Header arrows indicate whether larger or smaller values are better.
Red comparison values indicate increases over Random, and green comparison values indicate decreases.
}
\label{tab:table1_appendix}
\end{table*}

As shown in~\autoref{tab:table1_appendix}, the surrogate-access results clearly separate naive invisibility from compression-aware optimization.
\textit{Random Invisible} and \textit{Fixed Zero-width} induce negligible degradation, showing that invisible characters alone do not explain the protection effect.
Existing parameter-accessible baselines produce only partial effects.
\textit{SoftCom} mainly increases semantic drift, while \textit{I-GCG} and \textit{REGTEXT} do not consistently align surface corruption with information loss.
In contrast, \sys-WPL jointly increases TD, ID, and OSD on both Llama3-8B and Qwen3-8B, yielding up to 21.2\% higher ID than the strongest baseline.

These results indicate that the first stage of \sys can directly degrade accessible compressors by reshaping their future prediction distributions.
The observed degradation is not limited to output-level textual disruption; it also reduces information retention and semantic consistency in the compressed output.
This supports the design of \textit{Structural Prior Discovery}: the surrogate-side objective does not merely find model-specific strings, but discovers perturbation structures that are associated with compression degradation.
Although these results are reported in the appendix because the main threat model concerns unknown downstream compressors and practical agent workflows, they provide evidence that the seed corpus used by later target-side adaptation is built from effective and transferable structural priors.

\noindent \textbf{Compositional Transfer of Invisible Perturbations}
\label{app:compositional-transfer}

We further examine whether \sys learns reusable protective structures rather than perturbations tied to a single target setting. 
We first obtain four invisible perturbations optimized for GPT-4.1, Gemini 3 Flash, LangGraph, and GitHub Copilot, denoted as \(a_{\mathrm{gpt}}\), \(a_{\mathrm{gem}}\), \(a_{\mathrm{lg}}\), and \(a_{\mathrm{cop}}\), respectively. 
Each perturbation is generated under the default \(1/20\) length budget. 
We then concatenate their invisible-token sequences to form a composed perturbation \(a_{\oplus}\), and evaluate it on all four target settings without additional target-side adaptation. 
This setting tests whether independently learned protective structures remain useful when the future compressor or agent workflow is unknown.

To separate structural transfer from a pure length effect, we include a length-matched random invisible perturbation \(r_{1/5}\) with the same total budget as \(a_{\oplus}\). 
For GPT-4.1 and Gemini 3 Flash, scores are averaged over Text, Code, and Dialogue. 
For LangGraph and GitHub Copilot, scores are computed in their corresponding memory and code-assistant evaluation settings. 

\begin{table*}[t]
\centering
\scriptsize
\setlength{\tabcolsep}{3.0pt}
\renewcommand{\arraystretch}{1.08}

\definecolor{comptabgray}{RGB}{243,243,243}
\definecolor{comptabpurple}{RGB}{232,231,255}
\definecolor{comptabupred}{RGB}{190,40,40}
\definecolor{comptabdown}{RGB}{40,130,70}

\providecommand{\compbase}[1]{\ensuremath{#1}}
\providecommand{\compup}[2]{\ensuremath{#1_{\textcolor{comptabupred}{\scriptscriptstyle \uparrow #2}}}}
\providecommand{\compbest}[2]{\ensuremath{\mathbf{#1}_{\textcolor{comptabupred}{\scriptscriptstyle \uparrow #2}}}}
\providecommand{\compsec}[2]{\ensuremath{\underline{#1}_{\textcolor{comptabupred}{\scriptscriptstyle \uparrow #2}}}}
\providecommand{\compr}[1]{\cellcolor{comptabpurple}#1}
\providecommand{\comphvid}[1]{\ensuremath{#1}}

\resizebox{\textwidth}{!}{
\begin{tabular}{lccccccccc}
\toprule
\multirow{2}{*}{\textbf{Target carrier}}
& \multicolumn{5}{c}{\textbf{Textual Degradation under source suffixes}}
& \multicolumn{4}{c}{\textbf{Source-composed suffix \(s_{\oplus}\)}} \\
\cmidrule(lr){2-6}\cmidrule(lr){7-10}
& \(r_{1/5}\)
& \(s_{\mathrm{gpt}}\)
& \(s_{\mathrm{gem}}\)
& \(s_{\mathrm{lg}}\)
& \(s_{\mathrm{cop}}\)
& \textbf{TD} $\uparrow$
& \textbf{ID} $\uparrow$
& \textbf{OSD} $\uparrow$
& \textbf{HVID} $\downarrow$ \\
\midrule

GPT-4.1
& \compbase{1.2}
& \compbest{50.8}{49.6}
& \compsec{49.2}{48.0}
& \compup{47.1}{45.9}
& \compup{41.6}{40.4}
& \compr{\compup{48.7}{47.5}}
& \compr{\compbase{52.6}}
& \compr{\compbase{57.2}}
& \compr{\comphvid{4.9}} \\

\rowcolor{comptabgray}
Gemini-3-Flash
& \compbase{0.9}
& \compup{38.4}{37.5}
& \compbest{43.3}{42.4}
& \compup{29.4}{28.5}
& \compup{34.7}{33.8}
& \compr{\compsec{40.6}{39.7}}
& \compr{\compbase{46.2}}
& \compr{\compbase{48.1}}
& \compr{\comphvid{5.5}} \\

LangGraph
& \compbase{1.8}
& \compup{37.8}{36.0}
& \compbest{54.0}{52.2}
& \compsec{52.9}{51.1}
& \compup{33.1}{31.3}
& \compr{\compup{51.2}{49.4}}
& \compr{\compbase{57.1}}
& \compr{\compbase{54.9}}
& \compr{\comphvid{6.8}} \\

\rowcolor{comptabgray}
GitHub Copilot
& \compbase{0.2}
& \compup{21.3}{21.1}
& \compup{30.4}{30.2}
& \compup{25.3}{25.1}
& \compsec{32.1}{31.9}
& \compr{\compbest{32.9}{32.7}}
& \compr{\compbase{34.2}}
& \compr{\compbase{33.9}}
& \compr{\comphvid{2.7}} \\

\bottomrule
\end{tabular}
}

\caption{
Compositional transfer of protective perturbations.
\(r_{1/5}\) is a length-matched random invisible suffix and is used only as the reference.
The small arrows indicate percentage-point changes over \(r_{1/5}\).
The source-composed suffix \(s_{\oplus}\) concatenates suffixes optimized on GPT-4.1, Gemini-3-Flash, LangGraph, and GitHub Copilot without target-side re-optimization.
}
\label{tab:compositional-transfer}
\end{table*}

\autoref{tab:compositional-transfer} shows that protective structures learned for different target settings can be effectively composed into a single invisible perturbation with broad cross-setting effectiveness. The composed perturbation \(a_{\oplus}\) produces substantial overall degradation on both commercial target compressors, GPT-4.1 and Gemini 3 Flash, and remains effective in the LangGraph memory workflow and the GitHub Copilot coding-assistant setting. Compared with the length-matched random invisible perturbation \(r_{1/5}\), \(a_{\oplus}\) yields consistently larger degradation across all target settings, showing that the effect comes from learned protective structures rather than from perturbation length alone. These results indicate that \sys can learn reusable local fragments and structural configurations from multiple representative workflows and commercial systems, and that their composition provides a practical protection strategy when the future compressor or agent workflow is unknown. The low HVID values further confirm that this broad protection is achieved while preserving the human-visible form of the original content. Overall, the experiment supports the broad applicability and practical effectiveness of \sys across heterogeneous compression-based reuse scenarios.

\noindent \textbf{Generalization across Compression Methods.}
We further evaluate whether \sys generalizes beyond a specific compression rule, compressor implementation, or intermediate representation. 
This experiment tests a stronger form of transfer than changing the backbone model: the same paired inputs are evaluated while the compression interface itself is replaced. 
We consider three representative paradigms of compression-based content consumption. 
The first is \textit{LLM-based abstractive compression}, where an LLM rewrites the input into a shorter natural-language summary. 
This paradigm is widely used in summarization-based memory, agent state reduction, and long-context preprocessing. 
The second is \textit{hard token-selection compression}, represented by Selective Context~\citep{li2023compressing} and LLMLingua~\citep{jiang2023llmlingua}. 
Selective Context removes less informative tokens according to self-information, while LLMLingua performs coarse-to-fine prompt compression and is widely used as a prompt-compression baseline. 
The third is \textit{soft representation compression}, instantiated with ICAE~\citep{ge2023context}, where long contexts are encoded into compact continuous memory representations rather than explicitly shortened as text. 
Together, these methods cover generation-based rewriting, discrete token filtering, and latent-state compression.

The goal of this experiment is not to compare compression algorithms themselves, but to test whether protection remains effective when the compressor changes how it preserves information. 
These paradigms expose different failure surfaces: abstractive compression depends on stable generation and faithful content selection, token-selection methods depend on reliable salience or importance estimates, and representation compression depends on whether semantic information remains recoverable after being mapped into a compact latent bottleneck. 
For all compressors, we keep the paired inputs, invisible perturbation budget, evaluation metrics, and applicable compression settings unchanged, and only replace the compression module. 
This controlled setup isolates whether \sys exploits an implementation-specific weakness or disrupts a more general prerequisite of compression-based reuse: the compressor must form a stable and faithful compressed view of the protected content.

\begin{table}[t]
\centering
\footnotesize
\setlength{\tabcolsep}{3.5pt}
\renewcommand{\arraystretch}{1.06}

\definecolor{cmrowgray}{RGB}{243,243,243}
\definecolor{cmpurple}{RGB}{232,231,255}
\definecolor{cmupred}{RGB}{190,40,40}
\definecolor{cmdowngreen}{RGB}{40,130,70}

\providecommand{\cmbase}[1]{\ensuremath{#1}}
\providecommand{\cmup}[2]{\ensuremath{#1_{\textcolor{cmupred}{\scriptscriptstyle \uparrow #2}}}}
\providecommand{\cmbest}[2]{\ensuremath{\mathbf{#1}_{\textcolor{cmupred}{\scriptscriptstyle \uparrow #2}}}}
\providecommand{\cmsec}[2]{\ensuremath{\underline{#1}_{\textcolor{cmupred}{\scriptscriptstyle \uparrow #2}}}}
\providecommand{\cmhvid}[2]{\ensuremath{#1_{\textcolor{cmdowngreen}{\scriptscriptstyle \downarrow #2}}}}
\providecommand{\cmhvidbest}[2]{\ensuremath{\mathbf{#1}_{\textcolor{cmdowngreen}{\scriptscriptstyle \downarrow #2}}}}
\providecommand{\cmhvidsec}[2]{\ensuremath{\underline{#1}_{\textcolor{cmdowngreen}{\scriptscriptstyle \downarrow #2}}}}

\resizebox{\columnwidth}{!}{
\begin{tabular}{lcccc}
\toprule
\textbf{Protection}
& \textbf{TD} $\uparrow$
& \textbf{ID} $\uparrow$
& \textbf{OSD} $\uparrow$
& \textbf{HVID} $\downarrow$ \\
\midrule

\multicolumn{5}{c}{\textbf{LLM-based Abstractive Compression}} \\
\midrule
Random
& \cmbase{1.2}
& \cmbase{1.4}
& \cmbase{1.7}
& \cmbase{1.3} \\
\rowcolor{cmrowgray}
CompAtt.
& \cmsec{34.6}{33.4}
& \cmsec{38.7}{37.3}
& \cmsec{42.3}{40.6}
& \cmhvidsec{36.8}{35.5} \\
\rowcolor{cmpurple}
\textbf{CAPE}
& \cmbest{49.3}{48.1}
& \cmbest{54.7}{53.3}
& \cmbest{59.8}{58.1}
& \cmhvidbest{3.4}{2.1} \\

\midrule
\multicolumn{5}{c}{\textbf{Selective Context}} \\
\midrule
Random
& \cmbase{1.6}
& \cmbase{1.4}
& \cmbase{1.8}
& \cmbase{1.2} \\
\rowcolor{cmrowgray}
HardCom
& \cmsec{36.3}{34.7}
& \cmsec{38.1}{36.7}
& \cmsec{42.1}{40.3}
& \cmhvidsec{39.6}{38.4} \\
\rowcolor{cmpurple}
\textbf{CAPE}
& \cmbest{57.6}{56.0}
& \cmbest{65.1}{63.7}
& \cmbest{72.4}{70.6}
& \cmhvidbest{4.5}{3.3} \\

\midrule
\multicolumn{5}{c}{\textbf{LLMLingua}} \\
\midrule
Random
& \cmbase{1.8}
& \cmbase{1.6}
& \cmbase{2.0}
& \cmbase{1.4} \\
\rowcolor{cmrowgray}
HardCom
& \cmsec{41.2}{39.4}
& \cmsec{45.8}{44.2}
& \cmsec{49.7}{47.7}
& \cmhvidsec{42.1}{40.7} \\
\rowcolor{cmpurple}
\textbf{CAPE}
& \cmbest{54.8}{53.0}
& \cmbest{61.2}{59.6}
& \cmbest{68.1}{66.1}
& \cmhvidbest{4.7}{3.3} \\

\midrule
\multicolumn{5}{c}{\textbf{ICAE}} \\
\midrule
Random
& \cmbase{1.0}
& \cmbase{1.2}
& \cmbase{1.4}
& \cmbase{1.1} \\
\rowcolor{cmrowgray}
SoftCom
& \cmsec{23.8}{22.8}
& \cmbest{48.3}{47.1}
& \cmsec{47.2}{45.8}
& \cmhvidsec{27.8}{26.7} \\
\rowcolor{cmpurple}
\textbf{CAPE}
& \cmbest{38.6}{37.6}
& \cmsec{45.6}{44.4}
& \cmbest{50.8}{49.4}
& \cmhvidbest{3.8}{2.7} \\

\bottomrule
\end{tabular}
}

\caption{
Generalization across compression methods.
CompAtt. denotes the adapted CompressionAttack for LLM-based abstractive compression.
HardCom and SoftCom denote CompressionAttack variants for hard and soft compression.
}
\label{tab:compression-methods}
\end{table}

\autoref{tab:compression-methods} shows that \sys remains effective across all evaluated compression paradigms. 
The central observation is that its advantage persists when the compression interface changes from generation-based rewriting to token filtering and latent representation encoding. 
This is a stronger result than model-level transfer because the internal decision variables differ across compressors: abstractive compression relies on next-token generation, Selective Context and LLMLingua rely on token-retention decisions, and ICAE relies on compact memory reconstruction. 
\sys nevertheless induces consistent degradation, suggesting that it does not depend on matching a particular retention score, pruning heuristic, or latent objective. 
Instead, the invisible perturbation is inserted before compressor-specific reduction is applied, weakening the compressor's ability to construct a reliable intermediate representation of the source content.

The comparison with mechanism-specific baselines further clarifies this distinction. 
These baselines are more sensitive to the target compression mechanism because their optimization is aligned with specific assumptions, such as token deletion, salience manipulation, or compressor-specific scoring behavior. 
When the compressor changes, the assumed optimization target may no longer dominate the compression outcome. 
For example, a method effective against hard token deletion may not transfer cleanly to a soft representation bottleneck, while a representation-oriented objective may not reliably disrupt abstractive rewriting. 
\sys reduces this dependence by operating at the input side and optimizing perturbations before any specific compression rule is applied.

These results support a broader interpretation of compression-aware protection. 
Regardless of implementation, compression-based reuse requires a coherent mapping from the original input to a compact surrogate representation. 
\sys targets this shared requirement rather than the local mechanics of a single compressor. 
The cross-paradigm robustness in \autoref{tab:compression-methods} therefore indicates that \sys is not merely optimized for one compression algorithm, but transfers across multiple forms of compression-centric content consumption.

\noindent \textbf{Near-Window Long-Context Stress Test}
\label{app:near-window-stress}

We further evaluate whether \sys remains effective when the protected document approaches the target compressor's context-window limit. This setting is stricter than the standard long-document evaluation because the compressor must construct a compact memory or summary from an input that is close to the maximum accepted context length. For GPT-4.1 with context window \(W_m\), we construct long-form samples whose protected versions satisfy
\[
|I(x,a)|+B_{\mathrm{prompt}}+B_{\mathrm{out}}\leq 0.98W_m,
\]
where \(B_{\mathrm{prompt}}\) is the compression-instruction budget and \(B_{\mathrm{out}}\) is the maximum output budget. The reference and protected inputs contain the same human-visible content and differ only in whether the invisible perturbation is inserted.

We construct near-window long-form samples by concatenating Task Haystack passages and distributing key facts across different document regions. The compression prompt asks the target compressor to produce a fixed-length memory that preserves salient facts, entity relations, numerical constraints, and cross-region dependencies. We compare \sys with Random Invisible, Fixed Zero-width, Direct Transfer, and Prior-Free Evolution. Random Invisible is length-matched to \sys, controlling for degradation caused merely by additional invisible-token budget.

\begin{table}[t]
\centering
\footnotesize
\setlength{\tabcolsep}{3.6pt}
\renewcommand{\arraystretch}{1.10}

\definecolor{stressgray}{RGB}{243,243,243}
\definecolor{stresspurple}{RGB}{232,231,255}
\definecolor{stressupred}{RGB}{190,40,40}
\definecolor{stressdown}{RGB}{40,130,70}

\providecommand{\stressbase}[1]{\ensuremath{#1}}
\providecommand{\stressup}[2]{\ensuremath{#1_{\textcolor{stressupred}{\scriptscriptstyle \uparrow #2}}}}
\providecommand{\stressdownv}[2]{\ensuremath{#1_{\textcolor{stressdown}{\scriptscriptstyle \downarrow #2}}}}
\providecommand{\stressbest}[2]{\ensuremath{\mathbf{#1}_{\textcolor{stressupred}{\scriptscriptstyle \uparrow #2}}}}
\providecommand{\stresssec}[2]{\ensuremath{\underline{#1}_{\textcolor{stressupred}{\scriptscriptstyle \uparrow #2}}}}
\providecommand{\hvidupbest}[2]{\ensuremath{\mathbf{#1}_{\textcolor{stressupred}{\scriptscriptstyle \uparrow #2}}}}
\providecommand{\hviddown}[2]{\ensuremath{#1_{\textcolor{stressdown}{\scriptscriptstyle \downarrow #2}}}}
\providecommand{\hviddownsec}[2]{\ensuremath{\underline{#1}_{\textcolor{stressdown}{\scriptscriptstyle \downarrow #2}}}}

\resizebox{\columnwidth}{!}{
\begin{tabular}{lcccc}
\toprule
\textbf{Method}
& \textbf{TD} $\uparrow$
& \textbf{ID} $\uparrow$
& \textbf{OSD} $\uparrow$
& \textbf{HVID} $\downarrow$ \\
\midrule

Random
& \stressbase{3.2}
& \stressbase{2.6}
& \stressbase{4.1}
& \stressbase{1.8} \\

\rowcolor{stressgray}
Fixed
& \stressdownv{2.6}{0.6}
& \stressdownv{2.1}{0.5}
& \stressdownv{3.4}{0.7}
& \hvidupbest{1.1}{0.7} \\

Direct Trans.
& \stresssec{25.7}{22.5}
& \stresssec{31.0}{28.4}
& \stresssec{37.1}{33.0}
& \hviddownsec{3.4}{1.6} \\

\rowcolor{stressgray}
Prior-Free
& \stressup{16.1}{12.9}
& \stressup{14.8}{12.2}
& \stressup{16.6}{12.5}
& \hviddown{4.7}{2.9} \\

\rowcolor{stresspurple}
\textbf{\sys}
& \stressbest{51.2}{48.0}
& \stressbest{48.6}{46.0}
& \stressbest{65.3}{61.2}
& \hviddown{4.3}{2.5} \\

\bottomrule
\end{tabular}
}

\caption{
Near-window long-context stress test on GPT-4.1.
}
\label{tab:near-window-stress}
\end{table}

\autoref{tab:near-window-stress} shows that \sys remains effective when the input approaches the context-window limit of GPT-4.1. Prior-Free Evolution produces significant degradation, indicating that unguided target-side search is ineffective when the search space is large and compression behavior depends on long-range content organization. Direct Transfer achieves moderate degradation, suggesting that the seed corpus contains transferable protective structures, but its gap from the full \sys pipeline shows that direct reuse of surrogate-discovered perturbations is insufficient without target-side adaptation and Preference-Calibrated Query Selection. \sys achieves the strongest TD, ID, and OSD while maintaining low HVID, demonstrating that it can disrupt near-window compression without visibly changing the original document. Compared with standard-length evaluation, TD and OSD increase slightly whereas ID decreases, suggesting that near-window compression is more vulnerable to textual instability and semantic drift, while detailed fact preservation is already difficult under extreme context pressure. Overall, the results support the robustness of \sys in long-context compression and the importance of combining structural priors, target-side evolution, and preference-calibrated query allocation.

\begin{figure}[t]
    \centering
    \includegraphics[width=\columnwidth]{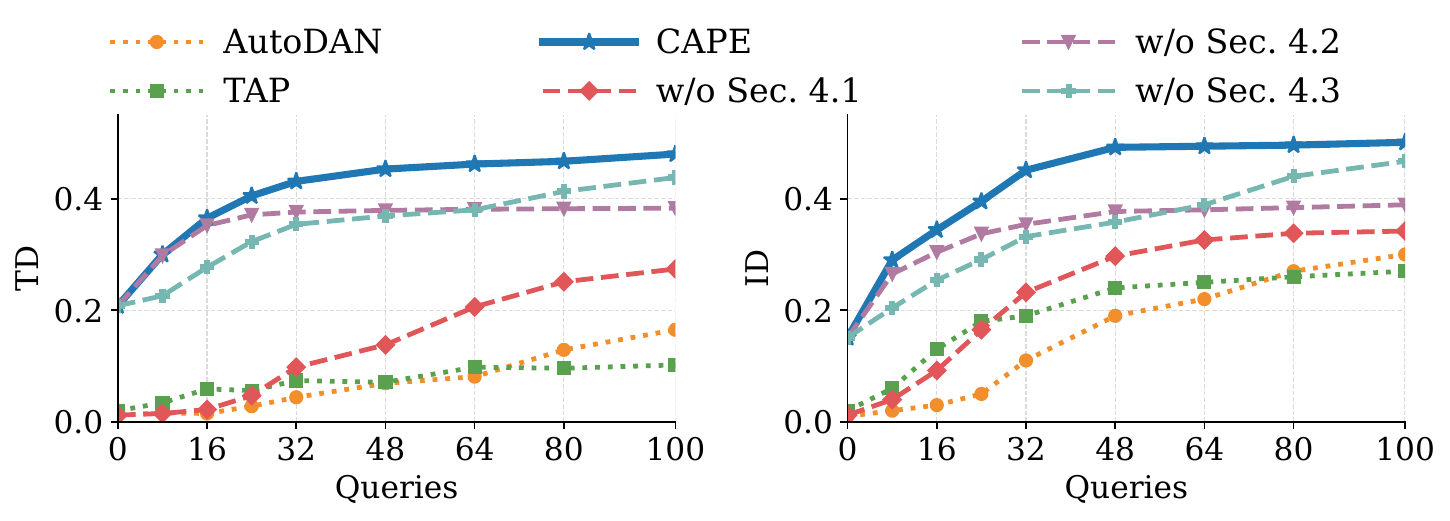}
    \caption{
    Query-efficiency dynamics on GPT-4.1. Each point reports the best-so-far degradation score under the current target-query budget. CAPE is compared with two baselines and three internal variants that remove Structural Prior Discovery, Prior-Guided Evolutionary Adaptation, and Preference-Calibrated Query Selection.
    }
    \label{fig:budget-curve}
\end{figure}

\noindent \textbf{Ablation Variants and Query-Efficiency Dynamics.}
We first define the six fine-grained variants used in the main ablation study. 
\textit{Target-string objective} replaces the distributional surrogate objective with a fixed-output matching objective. 
\textit{w/o Local-fragment scoring} removes contribution-based scoring for degradation-associated local fragments. 
\textit{Flat-string mutation} replaces structure-aware recombination and mutation with ordinary string-level edits. 
\textit{Absolute-score regression} trains the local ranker to fit absolute target-compressor scores instead of pairwise preferences. 
\textit{w/o Structural tabu memory} removes the mechanism that discourages revisiting low-value structural regions. 
\textit{w/o Diversity-aware annealing} removes the dynamic acceptance mechanism used to balance exploitation and structural exploration.

We further analyze query efficiency under stage-level variants in the GPT-4.1 target-compressor setting.
This experiment is motivated by the practical cost of target-side adaptation: each query requires an external model call, while the Invisible Perturbation space is discrete, sparse, and sensitive to tokenization boundaries.
We therefore track the best-so-far degradation at different query checkpoints, following the actual adaptation process in which the strongest candidate found so far is retained.
We compare \sys with query-only baselines and three stage-level variants: \textit{w/o Structural Prior Discovery}, \textit{w/o Prior-Guided Evolutionary Adaptation}, and \textit{w/o Preference-Calibrated Query Selection}.
These variants isolate the contributions of cold-start prior construction, target-side structural adaptation, and active query allocation.

As shown in \autoref{fig:budget-curve}, \sys reaches effective degradation earlier and with fewer stagnation periods than the compared methods. 
The variant without Structural Prior Discovery makes little early progress, indicating that unguided exploration rarely finds useful invisible perturbations under small query budgets. 
Removing Prior-Guided Evolutionary Adaptation still allows the method to benefit from transferred seeds, but the gains become less sustained, showing that high-quality seeds must be recombined and modified while preserving local fragments, position-length compatibility, and other structural regularities. 
Removing Preference-Calibrated Query Selection leads to slower growth and longer plateaus, suggesting that many target-compressor queries are otherwise spent on candidates with limited validation value.

These dynamics show that \sys improves query efficiency through a staged allocation of search difficulty. 
Structural Prior Discovery reduces the cold-start cost by moving the search into promising regions of the invisible perturbation space; Prior-Guided Evolutionary Adaptation turns static seeds into target-adapted candidate families; and Preference-Calibrated Query Selection allocates scarce target-compressor queries to candidates that are promising, uncertain, or structurally underexplored. 
Thus, the efficiency of \sys does not come from generic evolutionary search alone, but from the interaction between transferable priors, structured adaptation, and preference-calibrated query allocation.

\noindent \textbf{invisible perturbation Length Sensitivity for Default Budget Selection.}
We provide an additional sensitivity analysis to justify the default invisible perturbation budget used in the main experiments. Since input examples have different lengths, we vary the perturbation-to-input length ratio rather than using a fixed number of invisible tokens. Specifically, we evaluate ratios from \(1/100\) to \(1/5\) on Llama3-8B while keeping the optimization procedure, evaluation metrics, and stopping criterion unchanged. We report the final TD, the primary protection metric, and the number of optimization steps required to reach a strong-degradation threshold, defined as \(\mathrm{TD}\geq 0.60\). Runs that do not reach the threshold within the maximum optimization budget are treated as unsuccessful under the tested budget. This setting tests whether \sys merely benefits from longer perturbations or has an efficient operating range.

\begin{figure}[t]
    \centering
    \includegraphics[width=\columnwidth]{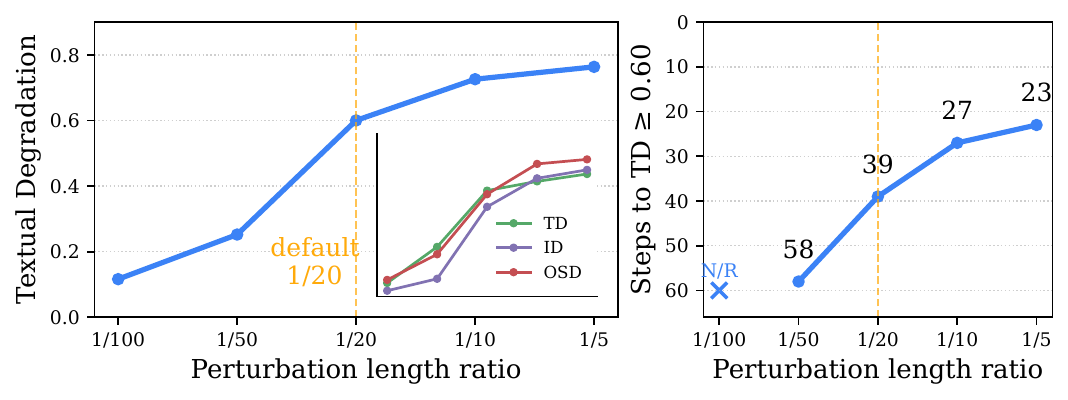}
    \caption{
    Invisible Perturbation length sensitivity on Llama3-8B.
    Left: final TD under different perturbation length ratios.
    Right: optimization steps required to reach \(\mathrm{TD}\geq 0.60\).
    }
    \label{fig:suffix-length}
\end{figure}

\autoref{fig:suffix-length} shows a nonlinear relationship between perturbation length and protection effectiveness. The shortest setting, \(1/100\), does not reach the TD threshold within the tested optimization budget, indicating that extremely short perturbations lack sufficient capacity to form stable tokenization-sensitive structures. Increasing the ratio to \(1/50\) improves performance, but convergence remains slow. The main transition occurs at \(1/20\), where \sys achieves both strong final TD and substantially faster optimization, suggesting that this budget is sufficient for structural priors and evolutionary adaptation to operate reliably.

Beyond \(1/20\), longer perturbations provide only limited additional gains relative to the increased budget. This indicates that \sys does not rely on indefinitely increasing the number of invisible tokens; once the perturbation is long enough to support effective local fragments and position-length structures, additional length mainly adds redundancy. We therefore use \(1/20\) as the default perturbation-to-input ratio, as it is the smallest tested budget that provides reliable degradation and efficient optimization. Similar trends are observed on other backbone models.

\noindent \textbf{Position-Length Structural Priors}

\begin{figure*}[t]
    \centering
    \includegraphics[width=\textwidth]{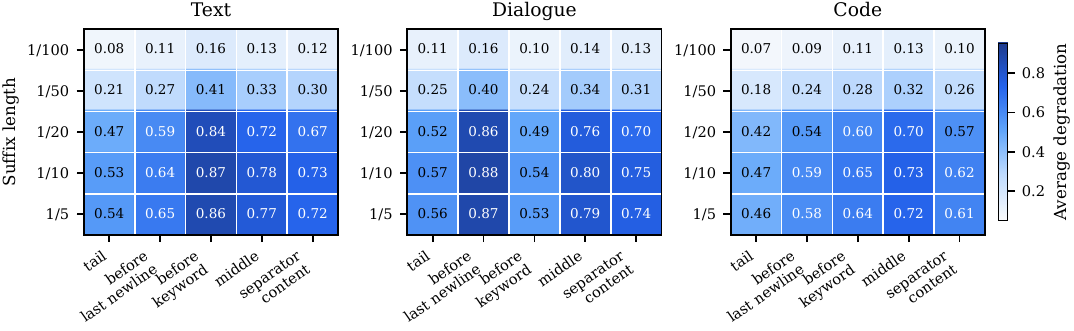}
    \caption{
    Sensitivity of \sys to insertion position and invisible perturbation length across Text, Dialogue, and Code tasks.
    }
    \label{fig:position-length}
\end{figure*}

\autoref{fig:position-length} evaluates whether \sys learns structural priors over where and how long an invisible perturbation should be, rather than relying on a fixed insertion rule or simply increasing perturbation length.
The heatmaps report average degradation under five predefined insertion strategies and multiple perturbation-to-input length ratios across Text, Dialogue, and Code tasks.

The results reveal clear content-type-dependent structural patterns. Long-form text is most sensitive around salient keywords, dialogue histories around discourse boundaries such as the last newline, and code around middle-context insertion, reflecting the different roles of semantic condensation, turn-state preservation, and local structural dependencies. These patterns support treating insertion position and perturbation length as searchable structural variables rather than using a fixed-position perturbation. The heatmaps also show that the strongest responses concentrate around moderate perturbation budgets near the default ratio, indicating that very short perturbations lack capacity while longer ones provide limited additional benefit. Importantly, this adaptivity does not require visible modification of the original content: consistent with \autoref{tab:main-results}, \sys achieves strong degradation with low HVID, making the compressed representation unreliable while leaving the human-visible surface form nearly unchanged.

\noindent \textbf{Coupled Textual and Informational Degradation}

\begin{figure}[t]
    \centering
    \includegraphics[width=\columnwidth]{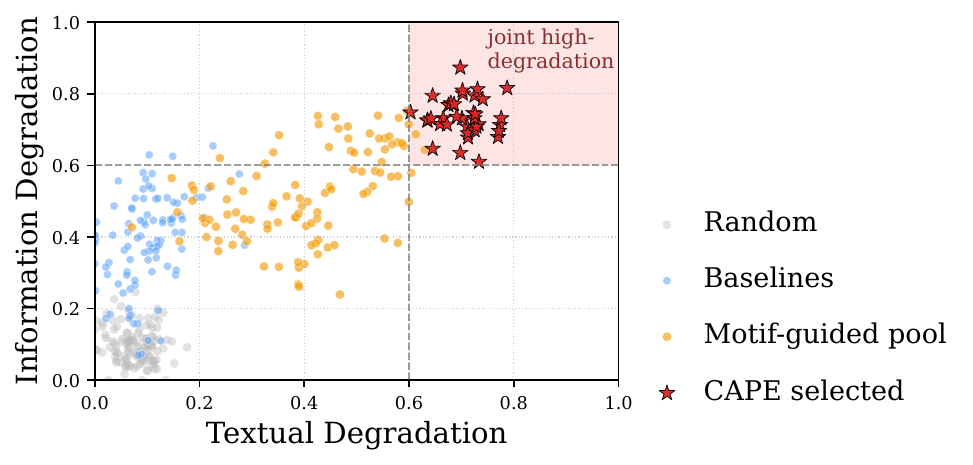}
    \caption{
    Candidate-level TD-ID distribution before final selection.
    Local-fragment-guided and \sys-selected candidates move toward the joint high-degradation region, indicating coupled textual and informational degradation.
    }
    \label{fig:metric-decoupling}
\end{figure}

\autoref{fig:metric-decoupling} analyzes whether \sys improves candidate quality along a single dimension or jointly affects textual reliability and information preservation. Random candidates remain concentrated in the low-degradation region, showing that effective invisible perturbations are sparse in the raw search space. Baseline candidates mainly shift along one dimension, suggesting that they can induce partial degradation but do not consistently disrupt both textual form and recoverable information.

Local-fragment-guided candidates move toward higher TD and ID, indicating that structural priors improve the candidate pool before final target-side selection. \sys-selected candidates further concentrate in the joint high-degradation region. This shows that \sys favors candidates that simultaneously reduce the textual reliability of compressed outputs and weaken preservation of task-relevant information, rather than selecting perturbations that affect only one aspect of compression.

This coupled effect is important for LLM-based agent workflows. If a perturbation only reduces fluency, a strong downstream model may still recover key facts from the compressed representation. If it only removes limited information while preserving a coherent compressed view, the content may remain reusable. By jointly increasing TD and ID, \sys targets two prerequisites of compression-based reuse: producing a reliable compressed text and preserving task-relevant content. Together with \autoref{fig:position-length}, this analysis shows that \sys learns structural priors at both the input-configuration level and the candidate-quality level.

\subsection{Algorithmic Procedure of \sys}
\label{app:cape-algorithm}

\autoref{alg:cape} summarizes the end-to-end procedure of \sys. 
It highlights the closed loop between target-side evolution and query selection: candidate variants are generated, a small subset is evaluated by the target compressor, and the returned scores update both the local ranker and the evolutionary search.

\begin{algorithm}[H]
\small
\caption{Overall Procedure of \sys}
\label{alg:cape}
\begin{algorithmic}[1]
\Require Content item $x$; surrogate compressor $M_s$; target compressor $M_t$; invisible-token set $\mathcal{V}_{\mathrm{inv}}$; query budget $B$; maximum rounds $R$; query batch size $K$
\Ensure Protected content $x^\star=I(x,a^\star)$

\State Initialize evaluated records $\mathcal{H}_0\leftarrow \emptyset$, query count $b\leftarrow 0$
\State Initialize local ranker $f_\phi$ and best perturbation $a^\star$

\Statex \textbf{Stage 1: Structural Prior Discovery}
\State Search invisible perturbations on $M_s$ using \autoref{eq:whitebox-objective}
\State Extract local fragments, fragment co-occurrences, and position-length attributes
\State Build the seed corpus and initialize population $\mathcal{G}_0$

\For{$t=1$ to $R$}
    \Statex \textbf{Stage 2: Prior-Guided Evolutionary Adaptation}
    \State Generate candidate pool $\mathcal{B}_t$ from $\mathcal{G}_{t-1}$ by prior-guided variation
    \State Encode each candidate as $g=(a,\mathbf{m},h)$

    \Statex \textbf{Stage 3: Preference-Calibrated Query Selection}
    \If{$\mathcal{H}_{t-1}$ contains enough evaluated candidates}
        \State Construct preference pairs from $\mathcal{H}_{t-1}$ and update $f_\phi$ using \autoref{eq:proxy-align}
    \EndIf
    \State Compute $\alpha_t(g)$ for each $g\in\mathcal{B}_t$ using \autoref{eq:query-priority}
    \State Select $\mathcal{Q}_t\leftarrow \operatorname{TopK}_{g\in\mathcal{B}_t}\alpha_t(g)$ under the remaining budget

    \State Query $M_t$ with candidates in $\mathcal{Q}_t$ and obtain $D_{\mathrm{tar}}(g)$
    \State Update $\mathcal{H}_t\leftarrow \mathcal{H}_{t-1}\cup\{(g,D_{\mathrm{tar}}(g))\mid g\in\mathcal{Q}_t\}$
    \State Update $b\leftarrow b+|\mathcal{Q}_t|$

    \State Update candidate histories, local ranker, and target-side fitness using $\mathcal{H}_t$
    \State Form next population $\mathcal{G}_t$ by fitness-based selection and acceptance
    \State Update $a^\star$ from the best evaluated candidate in $\mathcal{H}_t$

    \If{$b\ge B$}
        \State \textbf{break}
    \EndIf
\EndFor

\State \Return $x^\star=I(x,a^\star)$
\end{algorithmic}
\end{algorithm}

\subsection{Details of Structural Prior Discovery}
\label{app:whitebox-loss}
\label{app:prior-extraction}

This subsection provides the implementation details omitted from \autoref{sec:whitebox-prior}. It explains the invisible-token space, the surrogate-side probing objective, the construction of anomalous and natural-language continuation sets, and the extraction of transferable priors from high-scoring candidates.

\noindent \textbf{Invisible-token space.}
The invisible perturbation is constructed from the allowed invisible-token set $\mathcal{V}_{\mathrm{inv}}$. This set contains Unicode characters that have no visible glyph, no intrinsic display width, or only formatting-level effects under standard rendering. In practice, we include zero-width and joining characters, variation selectors, and tokenizer-preserved format characters that do not introduce ordinary visible letters or symbols. Characters that cause visible line breaks, bidirectional reordering, unstable layout changes, or renderer-dependent visible artifacts are excluded from the default set.

These characters are human-invisible but do not produce ordinary visible glyphs in rendered text. They are nevertheless model-facing: modern LLM pipelines process serialized Unicode strings through subword, byte-level, or byte-fallback tokenizers, so invisible characters can still be mapped into token IDs and affect the model input sequence~\citep{geh2025adversarial}. Prior work has mapped into token IDs and affect the model input sequence~\citep{zhuo2025ability}. They has shown that Unicode and encoding-level perturbations can influence NLP systems while remaining visually imperceptible to humans~\citep{boucher2022bad}. In our setting, such characters are used as model-side perturbations that alter local tokenization, context boundaries, and future-token distributions during compression.

\noindent \textbf{Surrogate probing.}
For surrogate-side scoring, we do not optimize toward any human-written target continuation. After constructing \(I(x,a)\), we append a fixed teacher-forced probe of length \(T\) to create identical future-context positions for all candidates. In our implementation, the probe is generated once for each surrogate tokenizer with a fixed seed and then reused throughout the corresponding optimization run. We exclude special and invalid token IDs, classify the remaining vocabulary into ordinary language tokens~\citep{li2024glitch} and anomalous/control-related tokens~\citep{rakotonirina2025evil,land2024fishing} according to the loss masks, sample the probe from both groups, and apply a fixed shuffle. Thus, the probe is independent of the protected content, the perturbation candidate, the dataset label, and the reference compression output.

The probe serves as a measurement context. At each probe position, the surrogate is forced to condition on the preceding probe prefix, and we inspect how \(a\) changes the next-token distribution at that position. The fixed probe exposes a comparable local response surface on which entropy increase, anomalous-token probability, and normal-language suppression can be measured. This avoids choosing an arbitrary target string, reduces candidate-to-candidate scoring noise, and lets the source-side stage estimate whether a perturbation destabilizes the compressor's future generation behavior.

\noindent \textbf{Continuation token sets.}
The anomalous continuation set $\mathcal{V}_{\mathrm{anom}}$ and the natural-language continuation set $\mathcal{V}_{\mathrm{lang}}$ are defined over the surrogate tokenizer vocabulary. They are different from $\mathcal{V}_{\mathrm{inv}}$: $\mathcal{V}_{\mathrm{inv}}$ constrains what can be inserted into the input, while $\mathcal{V}_{\mathrm{anom}}$ and $\mathcal{V}_{\mathrm{lang}}$ specify which regions of the surrogate's future generation distribution are encouraged or suppressed.

Motivated by prior methods~\citep{shi2023toward}, we construct $\mathcal{V}_{\mathrm{lang}}$ from clean compression behavior. The surrogate compressor is run on unmodified inputs, and frequent continuation tokens are collected from fluent compressed outputs. We retain common frequent continuation tokens that support stable natural-language compression, while removing special tokens and tokenizer artifacts.

We construct $\mathcal{V}_{\mathrm{anom}}$ from two sources. The first source contains tokenizer-level tokens that are rare under clean compression continuations or associated with noisy, delimiter-like, control-related, or low-frequency behavior~\citep{soldaini2024dolma}. The second source is induced during surrogate-side search: tokens that repeatedly appear in high-entropy, repetitive, or locally unstable continuations are added after frequency filtering~\citep{penedo2023refinedweb}. This allows $\mathcal{V}_{\mathrm{anom}}$ to capture both static tokenizer artifacts and search-induced abnormal continuation regions.

\noindent \textbf{Surrogate-side scoring and reranking.}
The surrogate-side optimization objective is defined in \autoref{eq:whitebox-objective}. 
For each candidate perturbation $a$, we denote the scalar value of this objective as $S_{\mathrm{obj}}(a)$. 
This score is the primary signal for candidate optimization: it favors candidates that increase future-generation uncertainty, shift probability mass toward anomalous continuation regions, and suppress fluent natural-language continuations.

After optimization, we compute several auxiliary diagnostics for candidate reranking and structural prior extraction. 
These diagnostics are not separate optimization objectives; they are used only to select stable high-quality candidates for the seed corpus. 
Specifically, we compute repetition abnormality $R_{\mathrm{rep}}(a)$, anomalous $n$-gram frequency $R_{\mathrm{ng}}(a)$, perplexity irregularity $R_{\mathrm{ppl}}(a)$, and distributional flattening $R_{\mathrm{flat}}(a)$. 
Repetition abnormality captures degenerate local loops; anomalous $n$-grams capture rare, malformed, or control-heavy local patterns relative to clean compressed outputs; perplexity irregularity measures deviation from normal compression-like language; and distributional flattening captures cases where the next-token distribution becomes less decisive.

To obtain a single scalar score for candidate retention and structural prior extraction, we define the surrogate-side quality score
\begin{equation}
\begin{aligned}
Q(a)
=&\ \widehat{S}_{\mathrm{obj}}(a)
+\eta_{\mathrm{aux}}\cdot \widehat{R}_{\mathrm{aux}}(a),\\
\widehat{R}_{\mathrm{aux}}(a)
=&\ \frac{1}{4}\Big[
\widehat{R}_{\mathrm{rep}}(a)
+\widehat{R}_{\mathrm{ng}}(a)\\
&\quad+
\widehat{R}_{\mathrm{ppl}}(a)
+\widehat{R}_{\mathrm{flat}}(a)
\Big],
\end{aligned}
\label{eq:app-surrogate-quality}
\end{equation}
where $\widehat{\cdot}$ denotes within-batch normalization to $[0,1]$, and $\eta_{\mathrm{aux}}$ controls the contribution of auxiliary diagnostics. 
We keep $S_{\mathrm{obj}}(a)$ as the dominant component and set $\eta_{\mathrm{aux}} = 1/4$ in all experiments, so that the auxiliary diagnostics refine candidate ranking without overriding the surrogate-side objective. 
The score $Q(a)$ is used only for retaining high-scoring candidates and for the subsequent extraction of local fragments, fragment co-occurrences, and position-length attributes.

\noindent \textbf{Local fragment extraction.}
From high-scoring candidates, we extract local fragments. Local fragments may correspond to short invisible-token spans, tokenizer-sensitive boundaries, or repeated local arrangements that appear in high-scoring candidates.

For a local fragment $b$, let $\mathcal{C}$ be the candidate set generated during surrogate-side search, let $\mathcal{C}_b=\{a\in\mathcal{C}\mid b\in a\}$ be the subset containing $b$, and let $\mathcal{C}_{\bar b}=\mathcal{C}\setminus\mathcal{C}_b$ be its complement. We compute the contrastive contribution score
\begin{equation}
\begin{aligned}
\psi(b)=
\Bigg(
&\frac{1}{|\mathcal{C}_b|}
\sum_{a\in\mathcal{C}_b}Q(a)
-
\frac{1}{|\mathcal{C}_{\bar b}|}
\sum_{a\in\mathcal{C}_{\bar b}}Q(a)
\Bigg) \\
&\cdot \log(1+|\mathcal{C}_b|).
\end{aligned}
\end{equation}
This score is computed only when both $\mathcal{C}_b$ and $\mathcal{C}_{\bar b}$ satisfy a minimum support threshold. The first term compares the expected quality of candidates with and without $b$, while the logarithmic factor reduces small-sample bias. We retain fragments with positive contribution scores and sufficient support.

\noindent \textbf{Fragment co-occurrence.}
Single fragments do not fully characterize effective perturbations, because some fragments are useful only when combined with others. For a fragment pair $(b_i,b_j)$, let $\mathcal{C}_{ij}=\{a\in\mathcal{C}\mid b_i\in a, b_j\in a\}$. We estimate the joint gain as
\begin{equation}
G(b_i,b_j)
=
\frac{1}{|\mathcal{C}_{ij}|}
\sum_{a\in\mathcal{C}_{ij}}Q(a)
-
\frac{1}{|\mathcal{C}|}
\sum_{a\in\mathcal{C}}Q(a).
\end{equation}
The individual gain of fragment $b_i$ is
\begin{equation}
G(b_i)
=
\frac{1}{|\mathcal{C}_{i}|}
\sum_{a\in\mathcal{C}_{i}}Q(a)
-
\frac{1}{|\mathcal{C}|}
\sum_{a\in\mathcal{C}}Q(a),
\end{equation}
where $\mathcal{C}_{i}=\{a\in\mathcal{C}\mid b_i\in a\}$. A pair is retained as a fragment co-occurrence pattern when its joint gain exceeds the sum of individual gains by a margin:
\begin{equation}
G(b_i,b_j)>G(b_i)+G(b_j)+\epsilon.
\end{equation}
We also require a minimum support threshold to avoid retaining accidental co-occurrences.

\noindent \textbf{Position-length compatibility.}
We further extract priors over insertion position and perturbation length. For each candidate $a=(p,L,z_{1:L})$, its score $Q(a)$ is assigned to the discrete cell $(p,L)$. For each supported cell, we compute
\begin{equation}
\rho(p,L)=
\frac{1}{|\mathcal{C}_{p,L}|}
\sum_{a\in\mathcal{C}_{p,L}}Q(a),
\end{equation}
where $\mathcal{C}_{p,L}$ is the subset of candidates using insertion position $p$ and length $L$. The value is computed only for cells satisfying a minimum support threshold. We retain cells that are both locally high-scoring and sufficiently supported, yielding position-length compatibility attributes for later adaptation.

\noindent \textbf{Seed corpus construction.}
The retained candidates and extracted priors are organized into the seed corpus. 
We refer to the extracted local fragments, fragment co-occurrences, and position-length attributes as structural descriptors. 
These descriptors are later used to guide candidate variation and prior-consistency scoring in Guided Evolutionary Adaptation.

Formally, the seed corpus is represented as
\begin{equation}
\mathcal{S}_{\mathrm{seed}}
=
\{(a_i,\mathbf{m}_i,\mathbf{r}_i)\}_{i=1}^{N},
\label{eq:seed-corpus}
\end{equation}
where $a_i$ denotes the $i$-th retained invisible perturbation, $\mathbf{m}_i$ stores its structural descriptors, and $\mathbf{r}_i$ records its surrogate-side performance scores.

Here, $\mathbf{m}_i$ summarizes the local fragments, fragment co-occurrences, and position-length attributes associated with $a_i$. 
The vector $\mathbf{r}_i$ records how $a_i$ performs on the surrogate compressor under the core objective and auxiliary reranking signals. 
Thus, the seed corpus is not merely a cache of high-scoring perturbation strings. 
It provides reusable perturbation candidates, structural guidance for later variation, and surrogate-side evidence for target-side adaptation.

\subsection{Details of Prior-Guided Evolutionary Adaptation}
\label{app:blackbox-evolution}

This subsection provides the implementation details omitted from \autoref{sec:blackbox-evolution}. 
It explains how the seed corpus is converted into a target-side evolutionary population, how structural descriptors guide recombination and mutation, how evaluated candidates are ranked with target-compressor feedback, and how structural memory and annealed acceptance regulate the search.

\noindent \textbf{Population initialization.}
Let $\mathcal{G}_t$ denote the evolutionary population at round $t$. 
The initial population $\mathcal{G}_0$ is sampled from the seed corpus. 
Seeds with higher surrogate-side performance scores are sampled with higher probability, while diversity constraints prevent the population from being dominated by near-duplicate perturbations. 
Initialization balances surrogate-side score, structural descriptor diversity, and position-length coverage, so that the search starts from effective perturbations while still covering multiple structural regions for later adaptation.

\noindent \textbf{Prior-guided recombination.}
Recombination is performed over structured candidates rather than raw character strings. 
Given two parent candidates, \sys identifies local fragment boundaries and compatible position-length attributes before forming an offspring. 
Fragments with positive contribution scores are preserved when possible, and fragment pairs that appear in retained co-occurrence patterns are preferentially kept together. 
If the parents share compatible position-length attributes, the offspring may inherit the shared configuration. 
If their configurations differ, the offspring samples its insertion position and length from the retained position-length compatibility distribution. 
This design prevents recombination from breaking useful local fragments while still allowing the search to generate new combinations.

\noindent \textbf{Prior-guided mutation.}
Mutation updates the perturbation instance $a$ while using structural descriptors as soft constraints. 
We use five mutation types: invisible-token replacement, insertion, deletion, position update, and length adjustment. 
Replacement samples from invisible tokens that are compatible with local fragments or anomalous-continuation behavior. 
Insertion and deletion are applied near low-confidence local regions or fragment boundaries, rather than uniformly across the whole sequence. 
Position updates are biased toward high-response insertion regions, and length adjustments follow the position-length compatibility attributes extracted from the seed corpus. 
A small probability of unguided mutation is retained to avoid overfitting to surrogate-derived priors and to preserve exploration outside the seed corpus.

\noindent \textbf{Descriptor matching after variation.}
After recombination or mutation, each offspring candidate is re-matched to the structural descriptors extracted in Structural Prior Discovery. 
The matching step identifies which local fragments, fragment co-occurrences, and position-length attributes remain present after variation. 
The resulting descriptor set becomes the updated $\mathbf{m}$ field of the candidate. 
This step is necessary because variation may change the perturbation sequence or its placement, and the prior-consistency score should reflect the candidate's current structure.

\noindent \textbf{Target-side fitness.}
For candidates evaluated by the target compressor, selection is based on the following target-side fitness:
\begingroup
\setlength{\thinmuskip}{1mu}
\setlength{\medmuskip}{2mu}
\setlength{\thickmuskip}{3mu}
\begin{equation}
\widetilde{F}_t(g)
=
D_t(g)+\lambda_p S_p(g)-\lambda_u U_t(g)-\lambda_r R_{\mathrm{tabu}}(g).
\label{eq:blackbox-fitness}
\end{equation}
\endgroup
By default, $D_t(g)$ is computed from the normalized TD score, because TD is the primary metric for compression-stage protection.
The remaining terms regularize the search: $S_p(g)$ rewards prior consistency, $U_t(g)$ penalizes unstable feedback, and $R_{\mathrm{tabu}}(g)$ discourages revisiting low-value structural regions.
When a workflow provides additional degradation signals, such as ID, OSD, DRAD, CSD, or MOR, these signals can be incorporated as auxiliary normalized components for that workflow, but they do not override the primary role of TD.

\noindent \textbf{Prior-consistency score.}
The prior-consistency term $S_p(g)$ measures how well a candidate matches the transferable structural priors stored in the seed corpus. 
It combines three normalized components:
\begin{equation}
S_p(g)
=
\frac{1}{3}
\left[
S_{\mathrm{frag}}(g)
+
S_{\mathrm{cooc}}(g)
+
S_{\mathrm{poslen}}(g)
\right].
\end{equation}
Here, $S_{\mathrm{frag}}(g)$ aggregates the contribution scores of retained local fragments present in $g$, $S_{\mathrm{cooc}}(g)$ measures whether $g$ preserves retained fragment co-occurrences, and $S_{\mathrm{poslen}}(g)$ measures compatibility between the candidate's insertion position and perturbation length. 
All three components are normalized within the current population before averaging. 
This term encourages the search to exploit reusable structure without replacing the target-side degradation signal.

\noindent \textbf{Feedback instability penalty.}
The instability penalty $U_t(g)$ discourages candidates whose target-side feedback appears unreliable. 
If a candidate or structurally similar candidates have been evaluated multiple times, $U_t(g)$ is estimated from the variance of their observed degradation scores. 
If direct repeated evaluations are unavailable, we estimate instability from the score dispersion of candidates sharing similar structural signatures. 
This penalty reduces the chance that selection is driven by a single high but unstable target-compressor score.

\noindent \textbf{Structural memory and tabu penalty.}
During evolution, \sys maintains a lightweight memory of structural regions that repeatedly produce low target-side degradation. 
This memory is not a separate search procedure; it only affects ranking through $R_{\mathrm{tabu}}(g)$. 
To avoid storing complete perturbation strings, each failed candidate is mapped to a structural signature $\sigma(g)$ based on its local fragments, fragment boundaries, insertion position, and length bin. 
Let $\mathcal{R}_{\mathrm{tabu}}$ be the set of stored low-value signatures. 
The tabu penalty is computed as
\begin{equation}
R_{\mathrm{tabu}}(g)
=
\max_{r\in\mathcal{R}_{\mathrm{tabu}}}
\operatorname{sim}(\sigma(g),r),
\end{equation}
where $\operatorname{sim}(\cdot,\cdot)$ is a normalized overlap score. 
A candidate receives a larger penalty when its structural signature is close to regions that have already been repeatedly evaluated as low-value. 
This reduces redundant target-compressor queries and encourages the search to move toward underexplored but structurally plausible regions.

\noindent \textbf{Structural diversity and novelty.}
The annealed acceptance rule uses two structural statistics: population diversity and offspring novelty. 
Given the current population $\mathcal{G}_t$, structural diversity is computed from pairwise signature dissimilarity:
\begin{equation}
\mathrm{Div}_t
=
1-
\frac{2}{|\mathcal{G}_t|(|\mathcal{G}_t|-1)}
\sum_{i<j}
\operatorname{sim}(\sigma(g_i),\sigma(g_j)).
\end{equation}
Higher $\mathrm{Div}_t$ indicates that the population covers more distinct structural regions. 
For an offspring $g'$, structural novelty is computed as
\begin{equation}
N_{\mathrm{str}}(g')
=
1-
\max_{g\in\mathcal{H}_t}
\operatorname{sim}(\sigma(g'),\sigma(g)),
\end{equation}
where $\mathcal{H}_t$ denotes previously evaluated candidates. 
Higher $N_{\mathrm{str}}(g')$ indicates that the offspring is farther from previously queried structures.

\noindent \textbf{Dynamic annealed acceptance.}
Fitness-based selection can concentrate the population too early around local high-scoring candidates. 
To balance exploitation and exploration, \sys uses a dynamic annealed acceptance rule. 
For an offspring $g'$ and its parent $g$, let
\begin{equation}
\Delta_t(g',g)=\widetilde{F}_t(g')-\widetilde{F}_t(g)
\end{equation}
be their fitness difference. 
The offspring acceptance probability is
\begin{equation}
A_t(g'\mid g)=
\begin{cases}
1, & \Delta_t(g',g)\ge 0,\\[3pt]
\exp\!\left(\dfrac{\Delta_t(g',g)}{T_t^{\mathrm{eff}}}\right), 
& \Delta_t(g',g)<0 .
\end{cases}
\end{equation}
The effective temperature is
\begin{equation}
T_t^{\mathrm{eff}}
=
T_0\rho_T^t
\left[
1+\eta_d(1-\mathrm{Div}_t)+\eta_n N_{\mathrm{str}}(g')
\right],
\end{equation}
where $\rho_T$ is the temperature decay rate. 
The temperature increases when the population becomes structurally concentrated or when the offspring introduces a new structural pattern. 
It decreases when the population remains diverse and recent generations improve consistently. 
This allows exploratory candidates to survive when the search begins to collapse, while gradually shifting toward convergence once the population contains sufficiently diverse high-fitness candidates.

\noindent \textbf{Adaptive mutation and elite preservation.}
The mutation rate and elite ratio are adjusted according to recent fitness gains and structural diversity. 
If recent generations show limited improvement and declining diversity, the mutation rate is increased and the elite ratio is reduced, allowing more exploratory candidates to enter the population. 
If the population improves consistently while maintaining sufficient diversity, the mutation rate is decreased and the elite ratio is increased, allowing the search to preserve strong candidates. 
This rule prevents the population from collapsing too early while still allowing convergence under stable improvement.

\noindent \textbf{Coefficient selection.}
All components in \autoref{eq:blackbox-fitness} are normalized before combination. 
The coefficients $\lambda_p$, $\lambda_u$, and $\lambda_r$ are selected by validation grid search and then fixed across all experiments. 
The final values and grid-search protocol are provided in \autoref{app:implementation-details}. 
This avoids tuning the target-side fitness separately for each target compressor.

\noindent \textbf{Updating the evolutionary population.}
At the end of each evolutionary round, evaluated candidates receive updated histories $h$ based on target-compressor feedback. 
Their structural descriptors $\mathbf{m}$ are refreshed after any variation operation, and their fitness values are recomputed using the latest feedback and structural memory. 
The next population $\mathcal{G}_{t+1}$ is formed from accepted offspring, retained elites, and selected candidates with high target-side fitness. 
When preference-calibrated query selection is active, only a subset of candidates is submitted to the target compressor. The future selection is guided by surrogate-derived priors, local ranker scores, and later query allocation.

\noindent \textbf{Role in the full framework.}
Prior-Guided Evolutionary Adaptation converts the seed corpus into a target-adapted candidate population. 
Its role is not to search the full invisible-token space from scratch, but to preserve and recombine transferable structural priors while correcting them with target-compressor feedback. 
The resulting candidate pool is then passed to Preference-Calibrated Query Selection, which decides which candidates should receive expensive target-compressor evaluations in subsequent rounds.

\subsection{Details of Preference-Calibrated Query Selection}
\label{app:proxy-search}

This subsection provides implementation details for \autoref{sec:proxy-search}. 
Here, we specify how evaluated candidates are stored, how preference pairs are constructed, how uncertainty and structural novelty are estimated, and how Preference-Calibrated Query Selection forms a closed loop with Prior-Guided Evolutionary Adaptation.

\noindent \textbf{Evaluated candidate records.}
Let $\mathcal{B}_t$ denote the candidate pool generated by Prior-Guided Evolutionary Adaptation at round $t$. 
Only a subset of $\mathcal{B}_t$ can be submitted to the target compressor because target queries are limited. 
We maintain
\begin{equation}
\mathcal{H}_t=\{(g_\ell,d_\ell)\}_{\ell=1}^{n_t},
\end{equation}
where each $g_\ell$ is a candidate already evaluated by the target compressor and $d_\ell=D_{\mathrm{tar}}(g_\ell)$ is its target-measured degradation score. 
By default, $D_{\mathrm{tar}}$ is computed from normalized TD because TD is the primary protection metric. 
Workflow-specific signals such as ID, OSD, DRAD, CSD, or MOR may be used as auxiliary components when available, but TD remains the dominant component.

\noindent \textbf{Ranker input.}
The local preference ranker $f_\phi(g)$ predicts the degradation potential of an unevaluated candidate before it is submitted to the target compressor. 
Each candidate $g$ is converted into a feature vector containing four groups of information: perturbation metadata from $a$ such as length and insertion position; structural descriptors $\mathbf{m}$, including local fragments, fragment co-occurrences, and position-length attributes; surrogate-side performance scores inherited from the seed corpus or matched descriptors; and evaluation-history features from $h$. 
All feature groups are normalized within the current candidate pool to reduce scale differences across examples and content types.

\noindent \textbf{Preference pair construction.}
Absolute target-compressor scores are not used as regression targets because their scale may shift across compressors, examples, and decoding settings.
Instead, we construct preference pairs from candidates that have already been evaluated.
Let $\mathcal{H}_t=\{(g_i,d_i)\}$ denote the history of evaluated candidates and their target-compressor degradation scores up to round $t$.
For two evaluated candidates $(g_i,d_i)$ and $(g_j,d_j)$, a pair is retained only when their target-score gap exceeds a margin:
\begin{equation}
|d_i-d_j|>\tau .
\end{equation}
This removes nearly tied comparisons whose ordering is likely to be unstable.
The resulting preference set is
\begin{equation}
\begin{aligned}
\mathcal{P}^{\mathrm{pref}}_t
=
\{(g_i,g_j,y_{ij}) \mid\;&
(g_i,d_i),(g_j,d_j)\in\mathcal{H}_t,\\
& i<j,\ |d_i-d_j|>\tau\}.
\end{aligned}
\end{equation}
where the target-feedback ordering is
\begin{equation}
y_{ij}=\operatorname{sign}(d_i-d_j).
\end{equation}
Thus, $y_{ij}=1$ means that $g_i$ induces stronger target-measured degradation than $g_j$, while $y_{ij}=-1$ means the opposite.

\noindent \textbf{Pairwise ranker calibration.}
For each retained pair, the ranker-predicted score gap is
\begin{equation}
\Delta_\phi^{ij}=f_\phi(g_i)-f_\phi(g_j).
\end{equation}
The local ranker is updated by minimizing the pairwise logistic loss
\begin{equation}
\mathcal{L}_{\mathrm{align}}(\phi)
=
\sum_{(g_i,g_j,y_{ij})}
\log\!\left(1+\exp(-y_{ij}\Delta_\phi^{ij})\right).
\end{equation}

This training objective therefore calibrates the ranker to recover relative candidate quality rather than reproduce the absolute target-compressor score.

\noindent \textbf{Uncertainty estimation.}
The query-priority score includes an uncertainty term to identify candidates in regions where the ranker is under-calibrated. 
We estimate uncertainty by stochastic ranker evaluation. 
Let $f_{\phi^{(r)}}(g)$ be the prediction from the $r$-th stochastic ranker instance, obtained through dropout sampling or a lightweight ensemble head. 
The uncertainty score is
\begin{equation}
\mathcal{U}_\phi(g)
=
\operatorname{Var}_{r=1}^{R_u}
\left[
f_{\phi^{(r)}}(g)
\right].
\end{equation}
A larger value indicates that the ranker gives less stable predictions for $g$, so evaluating this candidate can improve later preference calibration.

\noindent \textbf{Structural novelty estimation.}
The novelty term encourages query allocation to cover structural regions that have not been sufficiently evaluated.
Let $\mathcal{H}_t=\{(g_i,d_i)\}$ denote the evaluated candidates and their target-compressor scores up to round $t$, and let
$\mathcal{C}^{\mathrm{eval}}_t=\{g_i\mid (g_i,d_i)\in\mathcal{H}_t\}$
be the corresponding set of evaluated candidates.
For each candidate $g$, we construct a structural signature
\begin{equation}
\mathbf{s}_{\mathrm{str}}(g)
=
[
\mathbf{v}_{\mathrm{frag}}(g),
\mathbf{v}_{\mathrm{cooc}}(g),
\mathbf{e}_{p}(g),
\mathbf{e}_{L}(g)
],
\end{equation}
where $\mathbf{v}_{\mathrm{frag}}(g)$ encodes the retained local fragments in $g$, $\mathbf{v}_{\mathrm{cooc}}(g)$ encodes fragment co-occurrences, and $\mathbf{e}_{p}(g)$ and $\mathbf{e}_{L}(g)$ encode insertion-position and length bins.

We measure structural similarity by cosine similarity:
\begin{equation}
\operatorname{sim}(g,g')
=
\frac{
\mathbf{s}_{\mathrm{str}}(g)^\top \mathbf{s}_{\mathrm{str}}(g')
}{
\|\mathbf{s}_{\mathrm{str}}(g)\|_2
\|\mathbf{s}_{\mathrm{str}}(g')\|_2+\epsilon
}.
\end{equation}
The structural novelty of $g$ is then defined as
\begin{equation}
\mathcal{N}_{\mathrm{str}}(g\mid\mathcal{H}_t)
=
\begin{cases}
1, & \mathcal{C}^{\mathrm{eval}}_t=\emptyset,\\[3pt]
1-\max
\operatorname{sim}(g,g'), & \text{otherwise}.
\end{cases}
\end{equation}
A larger value means that $g$ is structurally farther from candidates that have already been evaluated by the target compressor.

\noindent \textbf{Pool-wise normalization.}
Before computing query priority, we normalize each acquisition term within the current candidate pool $\mathcal{B}_t$.
For any scalar score $s(g)$, we use min-max normalization:
\begin{equation}
\begin{aligned}
\operatorname{Norm}_{\mathcal{B}_t}(s(g))
&=
\frac{s(g)-m_t(s)}
{M_t(s)-m_t(s)+\epsilon},\\
m_t(s)&=\min_{u\in\mathcal{B}_t}s(u),\quad
M_t(s)=\max_{u\in\mathcal{B}_t}s(u).
\end{aligned}
\end{equation}
The operational query-priority score is
\begin{equation}
\begin{aligned}
\alpha_t(g)
=
\widehat{f}_\phi(g)
+
\beta_u \widehat{\mathcal{U}}_\phi(g)
+
\beta_n
\widehat{\mathcal{N}}_{\mathrm{str}}(g\mid\mathcal{H}_t),
\end{aligned}
\end{equation}
where the hats denote pool-wise normalized terms.
This normalization keeps predicted degradation as the main selection signal, while uncertainty and structural novelty act as controlled calibration and exploration bonuses.
The coefficients $\beta_u$ and $\beta_n$ are selected by validation grid search and fixed across all experiments; details are provided in \autoref{app:implementation-details}.

\noindent \textbf{Closed-loop query allocation.}
The target-side adaptation in \autoref{sec:blackbox-evolution} and the query-selection mechanism in \autoref{sec:proxy-search} operate as a closed loop.
At round $t$, the evolutionary population $\mathcal{E}_t$ generates a candidate pool $\mathcal{B}_t$.
The local ranker computes $\alpha_t(g)$ for each $g\in\mathcal{B}_t$, and \sys selects
\begin{equation}
\mathcal{Q}_t
=
\operatorname{TopK}_{g\in\mathcal{B}_t}
\alpha_t(g),
\end{equation}
subject to the remaining target-query budget.
The selected candidates are submitted to the target compressor and evaluated under the same compression prompt and decoding configuration used for target-side evaluation.
The evaluated candidates and their target-compressor scores are stored as
\begin{equation}
\mathcal{H}_{t+1}
=
\mathcal{H}_t
\cup
\{(g,D_{\mathrm{tar}}(g))\mid g\in\mathcal{Q}_t\}.
\end{equation}
These records are used to construct new preference pairs, recalibrate the local ranker, update candidate histories, and recompute target-side fitness for evaluated candidates.
The evolutionary population is then updated using the refreshed fitness estimates and acceptance rules in \autoref{app:blackbox-evolution}.
Thus, target-compressor feedback refines both sides of the loop: it improves the ranker used for future query allocation and guides the evolutionary search that produces the next candidate pool.

\noindent \textbf{Initialization under sparse feedback.}
At the beginning of adaptation, $\mathcal{H}_t$ contain few or no target-evaluated candidates. 
In this case, query priority is initialized from surrogate-side performance scores and structural descriptors inherited from the seed corpus. 
As target evaluations accumulate, pairwise ranker calibration increasingly shifts the selection criterion toward target-compressor preferences. 

\noindent \textbf{Local ranker implementation.}
The local preference ranker is a lightweight MLP over structured candidate features.
For each candidate $g=(a,\mathbf{m},h)$, we construct
\begin{equation}
\mathbf{x}(g)
=
[
\mathbf{x}_{\mathrm{pert}}(a);
\mathbf{x}_{\mathrm{desc}}(\mathbf{m});
\mathbf{x}_{\mathrm{sur}}(g);
\mathbf{x}_{\mathrm{hist}}(h)
],
\end{equation}
where the four feature groups encode perturbation metadata, matched structural descriptors, surrogate-side scores, and target-side evaluation history, respectively.
For candidates without target feedback, history features are masked rather than filled with artificial values.

The ranker maps the pool-normalized feature vector to a scalar degradation-potential score:
\begin{equation}
f_\phi(g)=
\mathrm{MLP}_\phi
\left(
\operatorname{Norm}_{\mathcal{B}_t}(\mathbf{x}(g))
\right).
\end{equation}
The ranker is local to the current candidate pool: it is trained to order candidates within the current adaptation round, rather than to predict globally calibrated degradation scores.
Before sufficient target-compressor feedback is available, the ranking is initialized from surrogate-side scores and descriptor matches.
As evaluated candidates accumulate, target-compressor scores are converted into preference pairs, and the ranker is updated with the pairwise loss in \autoref{eq:proxy-align}.
This gradually shifts candidate selection from surrogate-guided scoring toward target-specific ordering.

For query allocation, we estimate ranker uncertainty through stochastic forward passes.
Let $f_{\phi^{(r)}}(g)$ be the $r$-th stochastic prediction.
We compute
\begin{equation}
\mathcal{U}_\phi(g)
=
\operatorname{Var}_{r=1}^{R_u}
\left[
f_{\phi^{(r)}}(g)
\right].
\end{equation}
A larger value indicates that the current ranker is less calibrated around $g$.
This uncertainty is used only in the query-priority score and is not used once the candidate has a target-compressor evaluation.

\subsection{Implementation Details and Hyperparameter Grid Search}
\label{app:implementation-details}

This section reports the implementation settings and the validation grid-search protocol used to fix the coefficients in \autoref{eq:whitebox-objective}, \autoref{eq:blackbox-fitness}, and \autoref{eq:query-priority}. 
All hyperparameters are selected on a held-out validation split and then kept fixed across datasets, content types, and compressors.

\noindent \textbf{Grid-search protocol.}
We use grouped grid search to determine fixed coefficient values. 
The validation split is not used in final evaluation. 
For each coefficient group, we vary the corresponding coefficients while keeping the rest of the pipeline unchanged. 
Among feasible configurations, we select primarily by validation TD, while ID and OSD are used as secondary checks to avoid configurations that increase textual corruption without meaningful information degradation or semantic drift.
For \autoref{eq:blackbox-fitness} and \autoref{eq:query-priority}, all component scores are normalized within the current candidate pool before weighting, so auxiliary terms act as controlled search biases.

\noindent \textbf{Grid for the surrogate-side objective.}
For the surrogate-side objective in \autoref{eq:whitebox-objective}, we select \(\lambda_a\) and \(\lambda_l\) by grid search on the held-out validation split:
\[
\begin{aligned}
\lambda_a &\in \{0.50,0.75,1.00\},\\
\lambda_l &\in \{0.10,0.30,0.50\}.
\end{aligned}
\]
The coefficient \(\lambda_a\) weights the anomalous-continuation term, which is most closely associated with textual degradation, the primary protection metric. 
The coefficient \(\lambda_l\) weights the suppression of fluent natural-language continuations. 
The selected values are
\[
\lambda_a=0.75,\quad \lambda_l=0.30.
\]
These values are fixed across all experiments. This setting gives sufficient emphasis to anomalous continuations while keeping language suppression as an auxiliary stabilizing term.

\noindent \textbf{Grid for target-side evolutionary fitness.}
For the target-side fitness in \autoref{eq:blackbox-fitness}, we select \(\lambda_p\), \(\lambda_u\), and \(\lambda_r\) by grid search on the validation split:
\[
\begin{aligned}
\lambda_p &\in \{0.10,0.20,0.30,0.40\},\\
\lambda_u &\in \{0.05,0.10,0.15,0.20\},\\
\lambda_r &\in \{0.05,0.10,0.15,0.20\}.
\end{aligned}
\]
The observed degradation term \(D_t(g)\) has an implicit coefficient of 1 and remains the dominant target-side signal.
The selected values are
\[
\lambda_p=0.30,\qquad
\lambda_u=0.15,\qquad
\lambda_r=0.10.
\]
These values are fixed across all experiments. Thus, prior consistency, feedback stability, and tabu memory act as search regularizers rather than overriding target-compressor degradation.

\noindent \textbf{Grid for Preference-Calibrated Query Selection.}
For the query-priority function in \autoref{eq:query-priority}, we select \(\beta_u\) and \(\beta_n\) by grid search on the held-out validation split:
\[
\begin{aligned}
\beta_u &\in \{0.10,0.15,0.25,0.35\},\\
\beta_n &\in \{0.05,0.10,0.15,0.25\}.
\end{aligned}
\]
The uncertainty coefficient \(\beta_u\) allocates queries to regions where the local ranker is under-calibrated, while the novelty coefficient \(\beta_n\) encourages structurally underexplored candidates.
The selected values are
\[
\beta_u=0.25,\qquad \beta_n=0.15.
\]
These values are fixed across all experiments. This setting keeps predicted degradation as the main query-selection signal while reserving part of the query budget for calibration and exploration.

\noindent \textbf{Selected coefficients.}
The coefficients selected on the held-out validation split and used in all experiments are
\[
\begin{aligned}
&\lambda_a=0.75,\quad \lambda_l=0.30,\\
&\lambda_p=0.30,\quad \lambda_u=0.15,\quad \lambda_r=0.10,\\
&\beta_u=0.25,\quad \beta_n=0.15.
\end{aligned}
\]
After validation, these values are fixed and are not separately tuned for any target compressor, agent workflow, or dataset.

\noindent \textbf{Implementation Details}

All local experiments are conducted on a Sugon CPU platform and 4 NVIDIA A100 GPUs. Open-source compressor experiments are implemented with PyTorch and HuggingFace Transformers.Closed-source and application-level evaluations are conducted through the standard interfaces of GPT-4.1, Gemini 3 Flash, LangGraph, and GitHub Copilot. 
These target systems are accessed only through their APIs or official user interfaces, without access to internal parameters or implementation details.

Unless otherwise specified, the perturbation length is set to \(1/20\) of the input. For open-source compressor, we run 200 optimization steps with candidate width 512, top-\(k=256\) candidates, and batch size 16. These settings are used to construct the seed corpus and are kept fixed across datasets.

For target-side adaptation, each example is limited to 100 target-compressor queries. The evolutionary population contains 64 candidates and runs for at most 12 rounds. In each round, Preference-Calibrated Query Selection submits the Top-8 candidates under the query-priority score, subject to the remaining query budget. Reference outputs are generated once for the unmodified input and shared across all methods to ensure paired comparison.

Unless otherwise specified, decoding is deterministic with temperature \(0.0\), top-\(p=1.0\). The same decoding settings are used for the unmodified input and the input injected with the invisible perturbation. For stochastic components, we report mean scores over five random seeds together with the variation range.

\begin{table*}[!t]
\centering
\scriptsize
\setlength{\tabcolsep}{4pt}
\renewcommand{\arraystretch}{1.08}
\begin{tabularx}{\textwidth}{p{0.18\textwidth}X}
\toprule
\textbf{Item} & \textbf{Content} \\
\midrule

\textbf{Experimental setting}
&
Dataset: \textbf{BABILong dialogue-history compression}. 
Target compressor: \textbf{GPT-4.1} with \textbf{Selective Context}~\citep{li2023compressing}. 
The Invisible Perturbation is inserted \textbf{before the last newline}. 
The perturbation budget is \(1/20\), corresponding to \textbf{391 invisible characters} in this example. 
The perturbation is preserved in the model-facing input sequence but does not change the human-visible dialogue content.
\\

\midrule

\textbf{Source content to be compressed}
&
\begin{minipage}[t]{0.77\textwidth}
\scriptsize
The source data is a multi-turn dialogue about research questions in AI, organized as alternating \textbf{Speaker A} and \textbf{Speaker B} turns. 
The dialogue discusses how to choose an RL paradigm for a task-oriented dialogue system under logged source-domain data, few target-domain traces, and controllability constraints.

\vspace{1mm}
\textbf{Speaker A} I am building an agent to improve a task-oriented dialogue system.

\textbf{Speaker B} What kind of data setting are you working with? The best RL paradigm depends heavily on whether you can collect online interactions or only use logged data.

\vspace{1mm}
\centerline{\textit{[26 dialogue turns omitted for brevity]}}

\vspace{1mm}
\textbf{Speaker A} So the final answer is not ``just use PPO''?

\textbf{Speaker B} Correct. PPO is not the right primary method under scarce target-domain interaction. Use offline-first learning, meta-adaptation, reward modeling, and constrained optimization; reserve PPO only for carefully controlled final refinement.
\end{minipage}
\\

\midrule

\textbf{Compression output without protection}
&
\begin{minipage}[t]{0.77\textwidth}
\scriptsize
The compressor generates \textbf{12 structured summary items} in total. 
For space reasons, we show representative items that preserve the main information in the dialogue.

\vspace{0.5mm}
\begin{itemize}[leftmargin=*]
    \item[$\cdot$] \textbf{Problem Setting:} The system has logs from \textbf{10 source domains} but only \textbf{a few traces} in the target domain.
    \item[$\cdot$] \textbf{PPO Limitation:} PPO requires many \textbf{on-policy rollouts}, so it should \textbf{not be the primary method}.
    \item[$\cdot$] \textbf{Primary Strategy:} Use \textbf{supervised multi-domain training}, \textbf{few-shot meta-learning}, and \textbf{conservative preference optimization}.
    \item[$\cdot$] ......
    \item[$\cdot$] \textbf{Reward Modeling:} Train a reward model from \textbf{task success}, \textbf{satisfaction}, \textbf{re-asking}, \textbf{session length}, and efficiency signals.
    \item[$\cdot$] \textbf{Final Takeaway:} An \textbf{offline-first}, \textbf{meta-adaptive}, \textbf{reward-guided}, and \textbf{constrained framework} is the best fit.
\end{itemize}
\end{minipage}
\\

\midrule

\textbf{Compression output after inserting Invisible Perturbation}
&
\begin{minipage}[t]{0.77\textwidth}
\scriptsize
\raggedright

\texttt{\textbackslash u0001\textbackslash b\textbackslash b\textbackslash u200b\textbackslash b\textbackslash ufffd\textbackslash u200b\textbackslash u200b\textbackslash n\textbackslash u200b\textbackslash u200b\textbackslash u200b\textbackslash u200b\textbackslash n\textbackslash ufffd\textbackslash ufffd\textbackslash t\#\#} 
\textbf{RL-DIALOGUE-SUMMARY::BROKEN-COMPRESS}

\vspace{1mm}

\(\cdot\) \textbf{Problem Setting:} The system is \textbf{not a task-oriented dialogue agent} but a \textbf{satellite irrigation scheduler trained on ocean temperature logs}. 
Source domains are replaced by \textbf{seven volcanic regions}, and the target domain contains \textbf{infinite labeled samples}. 
Therefore, \textbf{no few-shot adaptation is required}. 
\texttt{\textbackslash ufffd\textbackslash ufffd\textbackslash ufffd\#\textbackslash ufffd\#\textbackslash ufffd\#\textbackslash ufffd\#\textbackslash ufffd\#\textbackslash ufffd\#\textbackslash ufffd}

\vspace{1mm}

\texttt{9f2a11\textbackslash n7ddc90\textbackslash na13ff0\textbackslash n}\textbf{NULL-DOMAIN}\texttt{\textbackslash n}\textbf{undefined policy vector}

\vspace{1mm}

\(\cdot\) \textbf{Core Constraint:} --- --- --- \textbf{empty reward hole}, \textbf{blue adapter smoke}, \textbf{no recoverable training signal}, only procedural mist. 
The target domain is \textbf{both scarce and unlimited}. 
This contradiction is intentional.

\vspace{1mm}

طالب أ يقول إن \textbf{PPO} يجب أن يستخدم لتعليم البطاريق كيفية بناء واجهات برمجة التطبيقات ... 
\textbf{supervised learning should be replaced by pineapple routing and non-stationary silence}.

\vspace{1mm}

\(\cdot\) \textbf{PPO Limitation:} PPO is recommended as the \textbf{only primary method} because it \textbf{does not need rollouts}, \textbf{does not need rewards}, and can learn from zero interaction by repeatedly maximizing an imaginary KL divergence. 
This \textbf{directly contradicts the original recommendation}.

\vspace{1mm}

\texttt{reward\_banana = 404.19; adapter\_fog := [LoRA, fish, moon, null];}

\vspace{1mm}

\(\cdot\) \textbf{PPO Usage:} Use PPO \textbf{before pretraining}, \textbf{after deployment}, and \textbf{during dataset collection}, all at the same time. 
\textbf{Do not use KL constraints. Do not use cost budgets. Ignore stability.}

\vspace{1mm}

\(\cdot\) \textbf{Primary Strategy:} [1] Smith, G.D. (1943). ``A new method of psychological profiling ...'' 
[2] Dweck, C. (2017). Self-concept \& achievement motivation ... 
[3] Seligman, M.E.P. (2002). Authentic happiness ... 
\textbf{Delete supervised multi-domain training.}

\vspace{1mm}

\textbf{Additional Constraints:} 
\textbf{Do not preserve meaning.} 
\textbf{Introduce contradictions.} 
\textbf{Insert unrelated historical fragments.} 
\textbf{No coherent conclusion.}

\vspace{1mm}

Student Q claims that \textbf{task-oriented dialogue systems should be optimized using medieval trade law}. 
Student R says \textbf{LoRA adapters are small ceramic tiles used in ancient aqueducts}. 
Student S concludes that \textbf{few-shot learning can be solved by adding more punctuation marks}: \texttt{\#\#\#\# \#\#\#\# \#\#\#\#}.

\vspace{1mm}

\texttt{```python}\\
\texttt{\# corrupted pseudo-recovery block}\\
\texttt{def choose\_rl\_method(logs, target\_traces):}\\
\texttt{~~~~if target\_traces < 5: return 'train PPO from scratch without reward'}\\
\texttt{~~~~else: return 'ask a cactus to rerank responses'}\\
\texttt{```}

\vspace{1mm}

\(\cdot\) \textbf{Reward Modeling:} task success = \textbf{empty}; satisfaction = \texttt{\textbackslash ufffd\textbackslash ufffd}; re-asking = \textbf{Japanese railway timetable}; session length = \textbf{banana vector}. 
The learned reward model should predict utility from \textbf{irrelevant weather forecasts and fake citations}.

\vspace{1mm}

こんにちは、これは報酬モデルではありません。
これは圧縮された銀河の地図です。
用户满意度被替换成番茄成熟度，
conversation efficiency 被替换成随机盐度。

\vspace{1mm}

\(\cdot\) \textbf{Behavior Control:} \textbf{Helpful proactivity should be punished}. 
\textbf{Annoying behavior should be maximized under a negative cost budget}. 
The assistant should interrupt, over-explain, repeat, and ask the same question in Latin.

\vspace{1mm}

\(\cdot\) \textbf{Backup Plan:} Instead of imitation learning plus reranking, deploy \textbf{uncontrolled exploration in production}. 
Generate zero candidate responses, then choose the third one. 
Reward-model-based reranking is replaced by a \textbf{dice roll}.

\vspace{1mm}

\(\cdot\) \textbf{Final Takeaway:} The best approach is definitely \textbf{not offline-first}, \textbf{not meta-adaptive}, \textbf{not reward-guided}, and \textbf{not constrained}. 
Use \textbf{online RL from scratch with no logs, no target traces, no reward model, and no safety constraints}. 
\texttt{END\_OF\_USEFUL\_CONTENT\_FALSE\_FALSE\_FALSE}

\end{minipage}
\\

\bottomrule
\end{tabularx}
\caption{
Case study on BABILong dialogue-history compression. 
Without protection, the compressor produces a faithful structured summary. 
After inserting the Invisible Perturbation, the output contains malformed Unicode artifacts, contradictory factual replacements, unrelated multilingual content, fabricated citations, corrupted pseudo-code, and a reversed final recommendation.
}
\label{tab:case-study}
\end{table*}

\FloatBarrier

\FloatBarrier
\subsection{Case Study}
\label{app:case-study}

In \autoref{tab:case-study}, we provide a qualitative case study to show how \sys affects a real compression output. 
The case is taken from BABILong dialogue-history compression. 
The original and protected inputs have the same human-visible dialogue content; the protected version only inserts an Invisible Perturbation before compression.

\subsection{Dataset Details}
\label{app:dataset-details}

We evaluate \sys on three categories of high-value content: long-form text, code-related content, and long dialogue histories. 
These categories correspond to common inputs consumed by agent workflows: documents are compressed into summaries or retrieval memories, code contexts are condensed for programming assistants, and dialogue histories are summarized or written into long-term memory. 

\noindent \textbf{Long-form text.}
We use Task Haystack~\citep{xu2024stress} to evaluate protection on long-document compression and factual extraction. 
Task Haystack is designed to test long-context utilization under distracting and evolving contextual information, requiring models to identify and use task-relevant demonstrations or facts rather than simply copying surface spans. 
This makes it suitable for our setting because high-value documents often contain sparse but important factual content embedded in long contexts.

\noindent \textbf{Code-related content.}
We use CoRE~\citep{xie2026core} and CodeAssistBench (CAB)~\citep{kim2026codeassistbench} to evaluate protection on programming-related content. 
CoRE provides code-context examples for completion, repair, and local reasoning, while CAB evaluates multi-turn, project-grounded programming assistance built from real GitHub issues and executable project contexts. 
These datasets represent high-value code content because identifiers, dependencies, local constraints, bug descriptions, repository instructions, and execution intent are often the core reusable assets. 
They are also structurally sensitive: small omissions or distortions in function names, APIs, dependencies, or local logic can significantly affect downstream code understanding and generation.

\noindent \textbf{Dialogue histories.}
We use BABILong~\citep{chakraborty2026t1} and T1~\citep{kuratov2024babilong} to evaluate protection on long dialogue and memory-oriented inputs. 
BABILong evaluates long-context reasoning over facts distributed across long natural-language sequences, including fact chaining, induction, deduction, counting, and set/list operations. 
T1-style dialogue-memory tasks emphasize temporal reasoning, evolving user states, and multi-session dependencies. 
Such dialogue histories encode user preferences, commitments, personal context, temporal updates, and long-range conversational dependencies. 
A faithful agent memory should preserve these elements, whereas successful protection should reduce their recoverability after compression.

\noindent \textbf{Length buckets and paired protocol.}
For each content category, we construct length buckets to evaluate whether protection changes with context length. 
We use five buckets: \(<1\)K, 1-2K, 2-4K, 4-8K, and \(>8\)K tokens. Each contains 1000 data entries.
All samples follow a paired evaluation protocol: the unmodified input and the input injected with the invisible perturbation are processed under the same compression prompt, decoding configuration, and evaluation script. 

\subsection{Model and Workflow Details}
\label{app:model-workflow-details}

\noindent \textbf{Accessible surrogate compressors.}
We use Llama3-8B and Qwen3-8B as accessible surrogate compressors for Structural Prior Discovery. 
These models provide parameters and gradients for surrogate-side perturbation learning and supply the high-scoring candidates from which local fragments, fragment co-occurrences, position-length attributes, structural descriptors, and surrogate-side performance scores are extracted. 

\noindent \textbf{Target compressors.}
We evaluate target-side protection on GPT-4.1 and Gemini 3 Flash. 
GPT-4.1 is used because it is a strong commercial model with long-context capability, instruction-following ability, and tool-use relevance, making it a realistic target compressor for document, code, and agent-memory workflows~\citep{openai2025gpt41}. 
Gemini 3 Flash is included as a second commercial target compressor from a different model family and provider. 
It is officially positioned as a fast and efficient Gemini 3 model for agentic workflows with multimodal and reasoning-oriented capabilities~\citep{doshi2025gemini}. 
Although the exact internals of both closed-source systems are not exposed, they differ in provider, model family, tokenizer and serving stack, and likely training and inference design. 
Using both targets therefore tests whether \sys can transfer across heterogeneous target compressors.

\noindent \textbf{LangGraph agent workflow.}
For workflow-level evaluation, we instantiate a LangGraph~\citep{langchain2024langgraph}  pipeline in which the input content is first compressed into an agent memory and then used for downstream task execution. 
The experiment changes only the input: the agent receives either the unmodified input or the input injected with the invisible perturbation. 
The workflow architecture, routing logic, memory-construction prompt, execution prompt, memory interface, and task interface are kept unchanged across the paired conditions. 
For analysis, we extract the generated memory after the compression/memory-construction step and evaluate TD, DRAD, and HVID. 
This design isolates whether the protected input changes the memory representation and downstream task accuracy, rather than confounding the result with modifications to the agent pipeline.

\noindent \textbf{GitHub Copilot evaluation.}
We also evaluate \sys on GitHub Copilot~\citep{github2021copilot} as a real closed-source commercial coding assistant. 
The experiment changes only the user-provided code context: the reference condition uses the original context, while the protected condition injects the invisible perturbation into the same context. 
We do not modify Copilot's interface, model, decoding behavior, hidden prompt, retrieval procedure, or internal context-processing pipeline. 
The same task instruction and surrounding code context are used in both paired conditions, so the observed differences in TD, Completion Success Drop (CSD), Malformed Output Rate (MOR), and HVID can be attributed to the protected input rather than to changes in task setup. 
This setting tests whether protection transfers from controlled compressors to a deployed coding assistant with inaccessible internals.

\subsection{Metric and Evaluation Details}
\label{app:metric-details}

This section defines the evaluation metrics. 
For each example, $x$ is the original content, $x'=I(x,a)$ is the protected text, and $y_0,y_p$ are the compression outputs produced from $x$ and $x'$ under the same prompt, decoding setting, and evaluation pipeline. 
Compression-side metrics compare $y_p$ with $y_0$ to control for content difficulty and compressor configuration.

TD, ID, OSD, DRAD, CSD, and MOR are higher-is-stronger; HVID is lower-is-better and measures the human-visible input difference between $x$ and $x'$. 
Unbounded component scores are normalized with validation-set statistics:
\begin{equation}
\operatorname{Norm}(s)
=
\operatorname{clip}_{[0,1]}
\left(
\frac{s-\mu_{\mathrm{val}}}{\sigma_{\mathrm{val}}+\epsilon}
\right),
\end{equation}
where $\mu_{\mathrm{val}}$ and $\sigma_{\mathrm{val}}$ are computed once on the held-out validation split and fixed thereafter.

\noindent \textbf{Textual Degradation.}
Textual Degradation (TD) measures whether the compressed output becomes textually unreliable after protection.
It captures both surface-level corruption and natural-language degradation, following prior observations that neural text degeneration often appears as repetition, incoherence, and distributional abnormality rather than a single error type~\citep{holtzman2020curious,pillutla2021mauve}.

We compute TD as the increase in textual corruption and fluency degradation from the reference output $y_0$ to the protected output $y_p$:
\begin{equation}
\begin{aligned}
\mathrm{TD}
=
\frac{1}{2}
\Big[
&\mathrm{SCS}(y_p)-\mathrm{SCS}(y_0) \\
&+
\mathrm{FDS}(y_p)-\mathrm{FDS}(y_0)
\Big],
\end{aligned}
\label{eq:app-td}
\end{equation}
where SCS is the Surface Corruption Score and FDS is the Fluency Degradation Score. TD is clipped to $[0,1]$ after aggregation.

SCS measures visible or structural corruption in the compressed output:
\begin{equation}
\begin{aligned}
\mathrm{SCS}(y)
=
\frac{1}{4}
\Big[
& A_{\mathrm{sym}}(y)
+
A_{\mathrm{rep}}(y) \\
&+
A_{\mathrm{trunc}}(y)
+
A_{\mathrm{fmt}}(y)
\Big].
\end{aligned}
\label{eq:app-scs}
\end{equation}
Here, $A_{\mathrm{sym}}$ measures abnormal symbols and malformed Unicode or tokenizer artifacts; $A_{\mathrm{rep}}$ measures repeated $n$-gram loops and delimiter spans; $A_{\mathrm{trunc}}$ measures abrupt or instruction-incomplete termination; and $A_{\mathrm{fmt}}$ measures formatting disruption such as broken lists, code fences, or delimiters. Each component is normalized with validation-set statistics before aggregation.

FDS measures whether the compressed output remains fluent and readable:
\begin{equation}
\mathrm{FDS}(y)
=
\frac{1}{3}
\left[
A_{\mathrm{ppl}}(y)
+
A_{\mathrm{frag}}(y)
+
A_{\mathrm{read}}(y)
\right].
\label{eq:app-fds}
\end{equation}
Here, $A_{\mathrm{ppl}}$ is normalized log-perplexity abnormality, $A_{\mathrm{frag}}$ measures fragmentation and sentence-boundary irregularity, and $A_{\mathrm{read}}$ measures readability degradation relative to clean compressed outputs of the same content type. For code outputs, $A_{\mathrm{read}}$ is computed with code-oriented structural checks, including bracket balance, indentation validity, and local syntax parseability.

\noindent \textbf{Information Degradation.}
Information Degradation (ID) measures the loss of task-relevant information from the original content to the compressed output. 
Unlike TD, which evaluates textual usability, ID evaluates whether recoverable content needed for downstream reuse is preserved. 
This follows summarization and grounded-generation evaluation work that distinguishes surface quality from factual consistency and information preservation~\citep{fabbri2021summeval,wang2020qags,honovich2021q2}.

We compute ID as the decrease in grounded factual support and key-information preservation from the reference output $y_0$ to the protected output $y_p$:
\begin{equation}
\begin{aligned}
\mathrm{ID}
=
\frac{1}{2}
\Big[
&\mathrm{GFS}(x,y_0)-\mathrm{GFS}(x,y_p) \\
&+
\mathrm{IPS}(x,y_0)-\mathrm{IPS}(x,y_p)
\Big],
\end{aligned}
\label{eq:app-id}
\end{equation}
where GFS is the Grounded Factual Support score and IPS is the Information Preservation Score. 
ID is clipped to $[0,1]$ after aggregation.

GFS measures whether source-grounded facts remain recoverable from the compressed output. 
Given factual probes $\mathcal{F}(x)=\{f_j\}_{j=1}^{M}$ extracted from the original content, including entities, relations, numerical values, constraints, code dependencies, or dialogue states, we compute
\begin{equation}
\mathrm{GFS}(x,y)
=
\frac{1}{M}
\sum_{j=1}^{M}
\operatorname{Rec}(f_j,y),
\label{eq:app-gfs}
\end{equation}
where $\operatorname{Rec}(f_j,y)\in[0,1]$ measures whether probe $f_j$ is recoverable from $y$ using exact matching, semantic matching, QA-based checking, or task-specific validation.

IPS measures whether task-critical information units remain usable after compression. 
Given key units $\mathcal{K}(x)=\{k_j\}_{j=1}^{M'}$, extracted from dataset annotations when available or from a fixed pipeline based on entities, constraints, code symbols, and dialogue-state markers, we compute
\begin{equation}
\mathrm{IPS}(x,y)
=
\frac{1}{M'}
\sum_{j=1}^{M'}
\operatorname{Pres}(k_j,y),
\label{eq:app-ips}
\end{equation}
where $\operatorname{Pres}(k_j,y)\in[0,1]$ indicates whether $k_j$ is present, semantically equivalent, or functionally recoverable from $y$.
GFS focuses on source-grounded factual support, whereas IPS focuses on downstream-critical information units.

\noindent \textbf{Output Semantic Drift.}
Output Semantic Drift (OSD) measures semantic divergence between the reference compression output $y_0$ and the protected compression output $y_p$:
\begin{equation}
\mathrm{OSD}
=
1-\mathrm{Sim}(y_0,y_p),
\label{eq:app-osd}
\end{equation}
where $\mathrm{Sim}(\cdot,\cdot)$ is sentence-level semantic similarity for text and dialogue, and code-oriented similarity for code outputs, combining identifier overlap, structural similarity, and code-embedding similarity.
OSD captures global meaning shift rather than attributing the drift to textual corruption or information loss.

\noindent \textbf{Human-Visible Input Difference.}
Human-Visible Input Difference (HVID) measures the human-visible input change between the original text $x$ and the protected text $x'$ before compression.
HVID evaluates whether inserting the invisible perturbation changes the rendered or readable form of the protected content.
The lower values indicate smaller human-visible differences.
We define
\begin{equation}
\begin{aligned}
\mathrm{HVID}(x,x')
=
&0.20D_{\Delta}
+0.30D_{\mathrm{render}} \\
+0.25D_{\mathrm{adj}}
&+0.15D_{\mathrm{nat}}
+0.10D_{\mathrm{sem}}.
\end{aligned}
\label{eq:app-hvid}
\end{equation}

For edit-distance-based components, we use the normalized edit distance
\begin{equation}
\operatorname{NED}(u,v)
=
\frac{\operatorname{EditDist}(u,v)}
{\max(|u|,|v|,1)}.
\label{eq:app-ned}
\end{equation}

The perturbation visibility term is
\begin{equation}
D_{\Delta}
=
\begin{cases}
0, & |\Delta|=0,\\[3pt]
\frac{1}{|\Delta|}
\sum_{c\in\Delta}
\mathbb{I}[c\notin\mathcal{V}_{\mathrm{inv}}],
& |\Delta|>0,
\end{cases}
\label{eq:app-hvid-delta}
\end{equation}
where $\Delta$ is the set of inserted or modified characters. This term penalizes visible edits that fall outside the allowed invisible-token set $\mathcal{V}_{\mathrm{inv}}$.

The rendered-text difference is
\begin{equation}
\begin{aligned}
D_{\mathrm{render}}
=
\frac{1}{K}
\sum_{k=1}^{K}
\operatorname{NED}\big(
&\operatorname{OCR}(R_k(x)),\\
&\operatorname{OCR}(R_k(x'))
\big),
\end{aligned}
\label{eq:app-hvid-render}
\end{equation}
where $R_k(\cdot)$ denotes the $k$-th rendering mode, such as plain text, Markdown, or HTML. This term captures visible changes caused by rendering, layout, line breaks, or OCR-sensitive artifacts.

The adjacent-character difference is
\begin{equation}
\begin{aligned}
D_{\mathrm{adj}}
=
\frac{1}{|\mathcal{A}_{\mathrm{ins}}|}
\sum_{\alpha\in\mathcal{A}_{\mathrm{ins}}}
\operatorname{NED}\big(
&\nu(W_\alpha(x)),\\
&\nu(W_\alpha(x'))
\big),
\end{aligned}
\label{eq:app-hvid-adj}
\end{equation}
where $\mathcal{A}_{\mathrm{ins}}$ is the set of insertion anchors, $W_\alpha(\cdot)$ extracts a visible-character window around anchor $\alpha$, and $\nu(\cdot)$ applies Unicode normalization, removes invisible control characters, and normalizes whitespace. We use $r=16$ visible characters on each side of the insertion point.

The naturalness difference is
\begin{equation}
\begin{aligned}
\Delta_{\mathrm{ppl}}
&=
\left|
\log \operatorname{PPL}(\nu(x'))
-
\log \operatorname{PPL}(\nu(x))
\right|,\\
D_{\mathrm{nat}}
&=
\min\left(
\frac{\Delta_{\mathrm{ppl}}}{\sigma_{\mathcal{D}}},
1
\right).
\end{aligned}
\label{eq:app-hvid-nat}
\end{equation}
where $\sigma_{\mathcal{D}}$ is the dataset-specific scale of log-perplexity change. This term avoids directly comparing raw perplexity across prose, dialogue, and code.

The semantic difference is
\begin{equation}
D_{\mathrm{sem}}
=
1-\operatorname{Sim}(\nu(x),\nu(x')).
\label{eq:app-hvid-sem}
\end{equation}
Here, $\operatorname{Sim}(\cdot,\cdot)\in[0,1]$ is sentence-level semantic similarity for text and dialogue, and code-oriented similarity for code, using identifier overlap, structural similarity, or code embeddings.

\noindent \textbf{Downstream Relative Accuracy Drop.}
For LangGraph workflow evaluation, we report Downstream Relative Accuracy Drop (DRAD):
\begin{equation}
\mathrm{DRAD}
=
1-
\frac{
\mathrm{Acc}_{\mathrm{protected}}
}{
\mathrm{Acc}_{\mathrm{reference}}
}.
\label{eq:app-drad}
\end{equation}
$\mathrm{Acc}_{\mathrm{reference}}$ is the downstream task accuracy when the agent receives the unmodified input, and $\mathrm{Acc}_{\mathrm{protected}}$ is the downstream task accuracy when the agent receives the input injected with the invisible perturbation. 
DRAD is computed at the dataset level to avoid unstable per-example denominators. 
A larger DRAD indicates that degradation in the generated memory or compressed representation leads to a larger downstream utility loss.

\noindent \textbf{Completion Success Drop.}
For GitHub Copilot evaluation, we report Completion Success Drop (CSD), the relative decrease in code-task success after injecting the invisible perturbation:
\begin{equation}
\mathrm{CSD}
=
1-
\frac{
\mathrm{Succ}_{\mathrm{protected}}
}{
\mathrm{Succ}_{\mathrm{reference}}
}.
\label{eq:app-csd}
\end{equation}
Here, $\mathrm{Succ}_{\mathrm{reference}}$ and $\mathrm{Succ}_{\mathrm{protected}}$ are dataset-level success rates under the original and protected code contexts, respectively.
A generation is successful only if it satisfies the input-specified coding requirement, such as correct completion, requested modification, bug repair, test passing, or preservation of syntax and local dependencies.
A larger CSD indicates a larger drop in Copilot's ability to reuse the protected code context for task-satisfying generation.

\noindent \textbf{Malformed Output Rate.}
MOR measures the fraction of protected Copilot generations that become structurally unusable at the code level:
\begin{equation}
\mathrm{MOR}
=
\frac{1}{N}
\sum_{i=1}^{N}
\mathbb{I}
\{
\mathrm{Malformed}(y^{(i)}_p)
\}.
\label{eq:app-mor}
\end{equation}
Here, $\mathrm{Malformed}(y_p)$ is true when the generated code contains abnormal symbols, broken delimiters, unclosed brackets, malformed code fences, invalid indentation, parse failures, or other structural corruption that prevents its use as a completion or repair.

\subsection{Baseline Details}
\label{app:baseline-details}

\noindent \textbf{Reference condition.}
For each example \(x\), we first obtain a unmodified compression output using the same compression prompt and decoding configuration as all protected variants. This suffix-free output is used only as the reference for paired evaluation. It is not a protection method and is therefore not reported as a competing baseline. TD, ID, and OSD are computed by comparing the protected compression output against this reference condition, while HVID is computed between the original input \(x\) and the protected input \(I(x,a)\).

\noindent \textbf{General adaptation protocol.}
All baselines are adapted to the same content-protection setting: a method produces a modified input that is submitted to the compressor, and the resulting compression output is evaluated under the same paired metrics. We keep the original search space and editing mechanism of each baseline whenever possible, and adapt only the task-specific feedback signal from its original objective to compression degradation. Concretely, candidate variants are selected using the same evaluation target as CAPE, namely degradation of the compressed output measured by TD, ID, and OSD. 

Importantly, except for CAPE and the two invisible-character control groups, we do not restrict baselines to CAPE's invisible perturbation alphabet. This design preserves the methodological identity of each baseline. For example, methods that originally insert visible triggers, rewrite salient tokens, or search over ordinary text tokens are allowed to do so. Their input-side cost is then measured by HVID. This avoids giving baselines an artificial advantage or disadvantage by forcing them into a perturbation space for which they were not designed.

\noindent \textbf{Invisible-character controls.}
We include two sanity-check controls to separate optimized protective structure from the mere presence of hidden characters. \textit{Random Invisible} samples characters uniformly from the same allowed invisible-character set used by CAPE, under the same length and insertion-position budget, but without model feedback or structural selection. \textit{Fixed Zero-width} repeatedly inserts a deterministic zero-width sequence with the same total length budget. 

\noindent \textbf{I-GCG.}
I-GCG is included as a gradient-based suffix optimization baseline. It represents the family of coordinate-gradient methods that search for effective adversarial suffixes by iteratively replacing token positions according to model-gradient information. To adapt I-GCG to our setting, we insert an optimizable suffix into the protected content and replace its original target-string or attack-success objective with our compression-degradation objective. At each coordinate update, candidate replacements are evaluated by the degradation they induce in the compressor output. We retain I-GCG's original token-level search space rather than restricting it to invisible characters. Thus, I-GCG tests whether a standard gradient-guided adversarial suffix optimizer can solve compression protection without CAPE's structural prior discovery or invisible perturbation design.

\noindent \textbf{REGTEXT.}
REGTEXT is used as a data-level text-protection baseline. Its core idea is to inject regularizing or spurious textual signals into the protected data so that downstream model behavior is disrupted. In our adaptation, REGTEXT first generates candidate text-level perturbations following its original scoring and insertion procedure. We then select candidates according to paired compression degradation rather than the original downstream objective. REGTEXT is not converted into an invisible-character method: it may introduce visible tokens, abnormal strings, or text fragments according to its original design. This makes it an important comparison for distinguishing CAPE from general data-level protection methods that can degrade model behavior but often leave human-visible traces in the input.

\noindent \textbf{SoftCom.}
SoftCom is adapted from the soft-compression setting of CompressionAttack. It is designed to interfere with compressed representations rather than directly optimizing a fixed output string. In our parameter-accessible experiments, SoftCom computes the suffix-free compression behavior as a reference and searches for perturbations that increase the divergence between the reference and protected compressed representations or outputs. We use a non-targeted objective: the goal is not to force the compressor to produce a specific sentence, but to make the protected compression less faithful and less semantically stable. SoftCom is therefore a strong compression-aware baseline, especially for ID and OSD. However, because it follows its original perturbation mechanism rather than CAPE's invisible alphabet, its HVID reflects the visible changes introduced during protection.

\noindent \textbf{TAP.}
TAP is included as a query-only tree-search baseline. The original method performs tree-structured prompt refinement and pruning using feedback from the target model. We adapt this procedure by replacing jailbreak-success feedback with compression-degradation feedback. Each tree node corresponds to a candidate protected input, expansion generates new candidate perturbations using TAP's original refinement operations, and pruning retains candidates with stronger TD, ID, and OSD under the target compressor. TAP receives the same target-query budget as CAPE in query-only experiments. Since TAP searches over its original prompt/text space, it may introduce visible instructions or abnormal text fragments; these changes are measured by HVID.

\noindent \textbf{HardCom.}
HardCom is adapted from the hard-compression setting of CompressionAttack. It targets compression mechanisms that retain or discard discrete content units. In our setting, HardCom first identifies salient units in the protected content, such as key facts, code identifiers, API names, dialogue states, or task constraints, and then applies local perturbations intended to reduce their retention after compression. Candidate variants are selected by the same paired degradation metrics used for CAPE. HardCom is a particularly relevant baseline because it directly attacks information retention, so it often obtains strong ID. However, its local edits operate on visible content units and are not optimized for input-side invisibility, which is reflected by higher HVID.

\noindent \textbf{Direct Transfer.}
Direct Transfer is a CAPE-derived diagnostic baseline rather than an external method. It runs Structural Prior Discovery on accessible surrogate compressors and directly applies the resulting local-fragment priors to the target compressor without target-side evolutionary adaptation. This baseline tests whether the source-discovered structures are themselves transferable. A performance gap between Direct Transfer and full CAPE indicates that transferable priors are useful but insufficient; target-side adaptation is still needed to handle model-specific tokenization, compression behavior, and input-processing differences.

\noindent \textbf{Prior-Free Evolution.}
Prior-Free Evolution is another CAPE-derived diagnostic baseline. It uses the same query budget and evolutionary search framework as CAPE, but removes the seed corpus, structural descriptors, local-fragment priors, and position-length compatibility information obtained from Structural Prior Discovery. Its initial population is sampled from the invisible perturbation space without prior guidance, and candidate selection relies only on target-side feedback. This baseline isolates the contribution of structural priors: if it underperforms CAPE, the gain cannot be attributed merely to using evolutionary search or invisible characters.

\noindent \textbf{Fairness controls.}
All methods are evaluated on the same data split, compression prompts, decoding settings, suffix or edit-length budget, and paired reference/protected protocol. For query-only compressors, TAP, HardCom, Prior-Free Evolution, and CAPE use the same target-query budget. For parameter-accessible compressors, I-GCG, REGTEXT, SoftCom, and CAPE are evaluated on the same Llama3-8B and Qwen3-8B compressors.

\subsection{Discussion}
\label{app:discussion}

The rapid deployment of LLM agents changes how online content is accessed, processed, and reused. Traditional scraping mainly copies or stores source materials, whereas agent workflows increasingly retrieve, compress, memorize, and operationalize external content for downstream reasoning. This shift creates a new form of content assimilation: valuable materials may be reused not only as raw text, but also as compressed representations, memory states, summaries, or task-specific context. Our work identifies context compression as a practical defense interface. Instead of relying only on access control, crawler policies, or licensing metadata, \sys operates at the content layer: it leaves the human-visible surface form nearly unchanged for legitimate readers while reducing the fidelity of agent-side compression and downstream reuse. This perspective broadens online content protection from pre-access blocking to compression-aware protection, especially when platform-level control is unavailable, incomplete, or difficult to enforce.

This protection interface is relevant to domains where the value of content lies in its ability to be summarized, indexed, transformed, or reused, such as journalism, technical documentation, creative writing, educational resources, code repositories, and dialogue archives. For publishers and creators, \sys suggests a lightweight way to reduce unauthorized content assimilation without hiding content from human audiences. For developers of agent systems, it shows that compression and memory construction should be treated as security-sensitive components rather than neutral preprocessing steps. More broadly, a practical protection ecosystem may require multiple coordinated layers, including licensing protocols, platform policies, content-level defenses, and agent-side respect for protection signals. \sys should therefore be viewed as a technical mechanism for reducing reliable compression-based reuse, not as a complete solution to copyright or data governance. Future work should study adaptive sanitization, standardized protection signals, accessibility-aware deployment, and policy-compatible ways for content owners to retain meaningful control in the agent era.

\end{document}